\documentclass[aps, pra, reprint, superscriptaddress, longbibliography]{revtex4-1}

\usepackage{natbib}
\usepackage[utf8]{inputenc}
\usepackage{amsmath}
\usepackage{amssymb}
\usepackage{hyperref}
\usepackage{graphicx}
\usepackage{epstopdf}
\usepackage{dcolumn}
\usepackage{soul}
\usepackage{bm}
\usepackage{soul}
\usepackage{xspace}
\usepackage{verbatim}
\usepackage{color}
\usepackage{xcolor}
\newcommand{\bra}[1]{\left\langle #1 \right|}
\newcommand{\ket}[1]{\left| #1 \right\rangle}

\hypersetup{colorlinks=true,linkcolor=blue,citecolor=blue, filecolor=blue,urlcolor=blue,breaklinks=true}

\begin{document}

\title{Heisenberg-Limited Spin-Mechanical Gravimetry}

\author{Victor Montenegro}
\email{vmontenegro@uestc.edu.cn}
\affiliation{Institute of Fundamental and Frontier Sciences, University of Electronic Science and Technology of China, Chengdu 611731, China}
\affiliation{Key Laboratory of Quantum Physics and Photonic Quantum Information, Ministry of Education, University of Electronic Science and
Technology of China, Chengdu 611731, China}

\date{\today}

\begin{abstract}
Precision measurements of gravitational acceleration, or gravimetry, enable the testing of physical theories and find numerous applications in geodesy and space exploration. By harnessing quantum effects, high-precision sensors can achieve sensitivity and accuracy far beyond their classical counterparts when using the same number of sensing resources. Therefore, developing gravimeters with quantum-enhanced sensitivity is essential for advancing theoretical and applied physics. While novel quantum gravimeters have already been proposed for this purpose, the ultimate sensing precision, known as the Heisenberg limit, remains largely elusive. Here, we demonstrate that the gravimetry precision of a conditional displacement spin-mechanical system increases quadratically with the number of spins: a Heisenberg-limited spin-mechanical gravimeter. In general, the gravitational parameter is dynamically encoded into the entire entangled spin-mechanical probe. However, at some specific times, the mechanical degree of freedom disentangles from the spin subsystem, transferring all the information about the gravitational acceleration to the spin subsystem. Hence, we prove that a feasible spin magnetization measurement can reveal the ultimate gravimetry precision at such disentangling times. We predict an absolute gravimetry uncertainty of $10^{-11}\text{m/s}^2$ to $10^{-6}\text{m/s}^2$, without relying on free-fall methodologies, ground-state cooling of the mechanical object, and robust against spin-mechanical coupling anisotropies.
\end{abstract}

\maketitle

\section{Introduction}

The measurement of gravitational acceleration, \textit{gravimetry}, is both theoretically and practically crucial. From a fundamental standpoint, gravimeters can aid in testing general relativity~\cite{ciufolini2016test, asenbaum2020atom}, for tackling relativistic quantum metrology problems~\cite{ahmadi2014quantum, ahmadi2014relativistic, bruschi2014quantum, kohlrus2017quantum, bravo2023gravitational, kohlrus2019quantum}, and for detecting deviations from Newtonian gravity~\cite{biswas2012towards}. From a practical viewpoint, gravimeters find extensive applications in geophysics~\cite{vancamp2017geophysics, pearson2018gravity, carbone2017added, van2017using, romaides2001comparison}, and thanks to satellite missions, it is now possible to shed light on planetary gravity fields and the complex dynamics between Earth's large masses of land, water, and atmospheric interactions~\cite{iess2018measurement, flechtner2021satellite, pail2014encyclopedia, ebbing2018earth, visser1999gravity}. 

Massive mechanical objects are excellent testbeds for gravimetry~\cite{rademacher2020quantum, marson1986g, zhong2022quantum} because their centers-of-mass are inherently subjected to gravitational acceleration and they straightforwardly couple to light and matter systems~\cite{treutlein2014hybrid}. While atomic interferometry techniques are predominantly employed for gravimetry~\cite{janvier2022compact, stray2022quantum, freier2016mobile, peters2001high, mcguirk2002sensitive, bidel2013compact, kritsotakis2018optimal, hu2013demonstration}, several other high-precision gravimeters have also been investigated. These include optomechanical systems~\cite{monteiro2017optical, monteiro2020force, armata2017quantum, feng2020enhancement, xiao2020optimal, qvarfort2021optimal, qvarfort2018gravimetry, johnsson2016macroscopic, geraci2015sensing}, ultracold Bose-Einstein condensate systems~\cite{abend2016atom, ke2017compact, phillips2022position, szigeti2020highprecision, gietka2019supersolid}, micro-electro-mechanical systems~\cite{middlemiss2016measurement, prasad202219day, mustafazade2020vibrating, tang2019high, krishnamoorthy2008inplane, liu2019passive}, diamagnetic-levitated micro-oscillator~\cite{leng2024measurement}, gravity-induced electric currents scheme~\cite{bruschi2023gravityinducedelectriccurrents}, and superconducting systems~\cite{merlet2021calibration}. Inspired by matter-wave interferometers~\cite{scala2013matterwave, wan2016free}, a gravimeter can be devised using a parametric spin-mechanical system~\cite{chen2018highprecision}, which can now be implemented with levitated optomechanics~\cite{rademacher2020quantum, hebestreit2018sensing, scala2013matterwave} or potentially in a more compact solid-state platform thanks to spin engineering~\cite{wei2006generation, Hendrickx2021four}. Among the advantages of this parametric coupling are (i) eliminating the need for matching resonance frequencies between the spin and mechanics~\cite{treutlein2014hybrid}, (ii) the high-fidelity control and read-out of the spin system~\cite{jiang2009repetitive, shields2015efficient, rabl2009strong}, and (iii) conditionally displacing the mechanical state based on known quantities (the coupling strength and its spin component)~\cite{scala2013matterwave, tufarelli2011oscillator, montenegro2014nonlinearity} and an unknown quantity (the gravitational acceleration)~\cite{scala2013matterwave, wan2016free}. Except for a few instances in gravimetry (e.g., see Ref.~\cite{gietka2019supersolid}) where the ultimate limit of precision has been predicted, known as Heisenberg limit, quantum-enhanced gravimetry at the Heisenberg level remains largely unexplored. We address this issue by exploring the precision limits of gravimetry in a parametric spin-mechanical system.

To achieve this goal, we employ quantum estimation theory~\cite{degen2017quantum, paris2009quantum}, which provides the theoretical framework for quantifying the precision of sensing unknown quantities. Specifically, for a given resource $N$ (such as the number of particles), the sensitivity of measuring an unknown parameter scales as ${\propto}N^\mu$~\cite{giovannetti2011advances, giovannetti2004quantum, giovannetti2006quantum, degen2017quantum, braun2018quantum}. Here, $\mu{=}1$ is the standard limit of precision achievable by classical strategies, $\mu{>}1$ signals quantum-enhanced sensitivity, and $\mu{=}2$ is generally known as the Heisenberg limit of precision. Thus, increasing the value of $\mu$ is a key figure of merit for enhancing sensitivity for a given resource $N$. Several questions must be addressed for parametric spin-mechanical gravimeters: (i) How can the ultimate gravimetry precision, the Heisenberg limit, be achieved? (ii) Does the Heisenberg limit hold when probing the system partially? (iii) Under which conditions can a feasible measurement attain the Heisenberg limit? (iv) How robust is the sensing procedure in the presence of noise?

In this work, we address all the above questions by demonstrating that: (i) the Heisenberg limit can be achieved by spins initialized in the Greenberger-Horne-Zeilinger (GHZ) state, namely the gravimetry uncertainty reduces quadratically with respect to the number of spins; (ii) the gravitational acceleration is solely encoded in the spin degree of freedom after each mechanical oscillator's cycle; hence, the Heisenberg limit can hold even if probing only the spin subsystem; (iii) at such a cycle, since the spin degree of freedom disentangles from the mechanics, one can achieve the Heisenberg limit by probing only the spin system using a feasible magnetization measurement basis; and (iv) the system shows robustness in the presence of coupling strength anisotropies and moderate decoherence at non-zero reservoir's temperature. We predict a sensitivity for moderate experimental parameters in the range of $10^{-11}\text{m/s}^2$ to $10^{-6}\text{m/s}^2$, without relying on free-fall methodologies or ground-state cooling of the mechanical object.

The rest of the paper is organized as follows: In Sec.~\ref{sec_information_metrics}, we provide a brief introduction to quantum parameter estimation. In Sec.~\ref{sec_themodel}, we describe the spin-mechanical gravimeter. In Sec.~\ref{sec_elementalgravimetry}, we present elementary gravitational scenarios. In Sec.~\ref{sec_SM_heisenberg}, we demonstrate that the gravimetry uncertainty decreases quadratically with the number of spin particles (Heisenberg limit) in the spin-mechanical probe at any time. In Sec.~\ref{sec_S_heisenberg}, we show that at specific times, the entire gravitational acceleration is transferred to the spin subsystem, achieving the Heisenberg limit exclusively within the spin subsystem. In Sec.~\ref{sec_CFI}, we determine the optimal spin measurement that achieves the Heisenberg limit. In Sec.~\ref{sec_robustness_anisotropies}, we investigate the robustness of the proposed gravimeter against spin-mechanical coupling anisotropies. In Sec.~\ref{sec_decoherence}, we analyze the gravimeter's performance under decoherence channels at finite temperature. In Sec.~\ref{sec_robustness_vicinity}, we study the robustness of the dynamical gravimeter scheme near the times when gravitational acceleration transfers to the spin subsystem. In Sec.~\ref{sec_feasibility}, we discuss the experimental feasibility of the proposed scheme. In Sec.~\ref{sec_predicted}, we calculate the predicted absolute sensitivity of the proposed spin-mechanical gravimeter. Finally, in Sec.~\ref{sec_conclusions}, we conclude this work.

\section{Information metrics}\label{sec_information_metrics}

This section introduces a concise overview of the key concepts and metrics in quantum estimation theory needed throughout this work. The uncertainty of an unknown parameter $g$ encoded in a quantum state $\rho_g$ is fundamentally lower bounded by the quantum Cram\'{e}r-Rao inequality~\cite{cramer1999mathematical, rao1992breakthroughs, paris2009quantum}:
\begin{equation}
    \mathrm{Var}[\hat{g}]\geq (\nu\mathcal{F})^{-1}\geq (\nu\mathcal{Q})^{-1},\label{eq_QCRB}
\end{equation}
where $\mathrm{Var}[\hat{g}]$ is the variance of $\hat{g}$, $\hat{g}$ is the estimator of $g$ (assumed here unbiased)~\cite{paris2009quantum}, $\nu$ is the number of experimental repetitions, $\mathcal{F}$ is the classical Fisher information (CFI), and $\mathcal{Q}$ is the quantum Fisher information (QFI). Needless to say, $\mathcal{F}(g){:=}\mathcal{F}$ and $\mathcal{Q}(g){:=}\mathcal{Q}$ depend on the unknown parameter $g$. However, for the sake of clarity, this dependence has been made implicit in their notation. The CFI is defined as $\mathcal{F}{=}\sum_\Upsilon p(\Upsilon|g)[\partial_g\mathrm{ln}p(\Upsilon|g)]^2$, where $\partial_g{:=}\partial/\partial g$ and $p(\Upsilon|g){=}\mathrm{Tr}[\hat{\Pi}_\Upsilon\rho_g]$ is the probability of measuring the quantum state $\rho_g$ with the positive-operator valued measure (POVM)~\cite{janos2021quantum} $\{\hat{\Pi}_\Upsilon\}$ with measurement outcome $\Upsilon$. Hence, the CFI yields the achievable precision for a given POVM. The POVM that maximizes the CFI is known as the QFI, $\mathcal{Q}{:=}\max\limits_{\{\hat{\Pi}_\Upsilon\}}\mathcal{F}$. Several expressions exist for the QFI~\cite{paris2009quantum}. In particular, for the entire spin-mechanical pure state, i.e., $\rho_g{=}\ket{\psi_g}\bra{\psi_g}$, the QFI simplifies as
\begin{equation}
    \mathcal{Q}_\mathrm{sm}{=}4\mathbb{R}\text{e}\left[\langle \partial_g \psi_g | \partial_g \psi_g \rangle{-}|\langle \partial_g \psi_g | \psi_g \rangle|^2\right]\label{eq_QFI_all},
\end{equation}
whereas for the mixed state of the reduced density matrix of the spin, the QFI is given by:
\begin{equation}
\mathcal{Q}_\mathrm{spin} = 2 \sum_{n,m} \frac{|\langle \lambda_m |\partial_g \rho_\mathrm{spin}| \lambda_n \rangle|^2}{\lambda_n + \lambda_m}, \hspace{0.5cm} \lambda_n + \lambda_m \neq 0,
\end{equation}
where $\rho_\mathrm{spin}{=}\sum_i \lambda_i |\lambda_i\rangle\langle\lambda_i|$ is represented in its spectral decomposition, with $\lambda_i$ and $|\lambda_i\rangle$ being the $i$th eigenvalue and eigenvector of $\rho_\mathrm{spin}$, respectively~\cite{paris2009quantum}. The QFI quantifies the sensing capability of the quantum probe $\rho_g$; thus, a higher QFI value indicates lower uncertainties in $g$, as shown in Eq.~\eqref{eq_QCRB}. Note that the QFI does not explicitly provide the optimal measurement basis, and one needs to build the POVM from the symmetric logarithmic derivative (SLD) operator $L_g$ satisfying $2\partial_g{\rho_g}{=}L_g\rho_g{+}\rho_gL_g$~\cite{paris2009quantum, liu2016quantum}. In addition, the ultimate precision limit is a consequence of the response of the quantum probe to infinitesimal changes in the vicinity of the unknown parameter~\cite{sidhu2020geometric, braunstein1994statistical}.

\section{The model}\label{sec_themodel}

We consider $N$ non-interacting spin-1/2 particles coupled to a single-mode harmonic oscillator via a conditional displacement Hamiltonian:
\begin{equation}
    \hat{H} = \frac{1}{2M}\hat{p}^2 + \frac{M\omega^2}{2}\hat{x}^2 - \sum_{i=1}^N \bar{k}_i \frac{\hat{\sigma}_i^z}{2} \hat{x} + \bar{g} M \hat{x} \cos\xi,\label{eq_hamiltonian}
\end{equation}
where $\hat{p}$ and $\hat{x}$, satisfying $[\hat{x}{,}\hat{p}]{=}i\hbar$, are the momentum and position operators of a mechanical object with mass $M$ and frequency $\omega$, respectively. $\hat{\sigma}_i^\beta$, $\beta{=}x{,}y{,}z$, is the Pauli operator for the $i$th spin, which interacts with the mechanical oscillator with strength $\bar{k}_i$. The final term describes the gravitational potential energy~\cite{qvarfort2018gravimetry, scala2013matterwave, armata2017quantum, chen2018highprecision}, where $\bar{g}$ is the gravitational acceleration, and $\xi$ is the angle between the $x-$direction and the free fall acceleration~\cite{qvarfort2018gravimetry, scala2013matterwave}. The Hamiltonian for a single spin interacting with a mechanical oscillator, in the absence of the gravitational term, has been extensively investigated~\cite{khosla2018displacemon, rabl2010quantum, rabl2009strong, rabl2010cooling, braccini2023large, rao2016heralded, montenegro2017macroscopic, tufarelli2012reconstructing, montenegro2018ground, tufarelli2011oscillator, scala2013matterwave, montenegro2014nonlinearity, yin2013large, spiller2006quantum, kumar2017magnetometrty}. By introducing $\hat{p}{=}i\sqrt{M\hbar\omega/2}(\hat{a}^\dagger{-}\hat{a})$ and $\hat{x}{=}\sqrt{\hbar/(2M\omega)}(\hat{a}^\dagger{+}\hat{a})$, one can write Eq.~\eqref{eq_hamiltonian} as follows:
\begin{equation}
\frac{\hat{H}}{\hbar \omega}{=}\hat{a}^\dagger \hat{a}{-}\hat{Z}(\bm{k}{,}g)\left(\hat{a}^\dagger{+}\hat{a}\right),\label{eq_hamiltonian_scaled}
\end{equation}
where $\hat{Z}(\bm{k}{,}g){:=}\left[\sum\limits_{i=1}^N k_i \frac{\hat{\sigma}_i^z}{2}{-}g\right]$ is a spin operator depending on the coupling strengths $\bm{k}{=}{\{k_i\}}$ and the unknown parameter $g$ to be estimated. Moreover, we have defined the quantities $\hbar k_i{:=}\bar{k}_i\sqrt{\hbar/(2M\omega^3)}$, $\hbar g{:=}\bar{g}\sqrt{M\hbar/(2\omega^3)}\cos\xi$. Without loss of generality, we will set $\xi{=}0$ from this point forward. To solve the quantum dynamics, it has been analytically derived the unitary temporal operator for Eq.~\eqref{eq_hamiltonian_scaled} as~\cite{montenegro2014nonlinearity}:
\begin{equation}
    \hat{U}(\tau){=}e^{ i \hat{Z}(\bm{k}{,}g)^2(\tau{-}\sin\tau)}e^{\hat{Z}(\bm{k}{,}g)(\eta(\tau)\hat{a}^\dagger{-}\eta^*(\tau)\hat{a})} e^{-i\hat{a}^\dagger\hat{a}\tau},\label{eq_unitary_operator}
\end{equation}
where $\eta(\tau){:=}1{-}e^{-i\tau},$ and $\tau{:=}\omega t$. Thanks to mechanical oscillators' current ground state cooling~\cite{oconnell2010quantum}, we will initialize the mechanical object in a coherent state $|\alpha\rangle$, $\alpha{\in}\mathbb{R}$. This is not a limitation, as the spin-mechanical gravimeter used here operates effectively even when the mechanical oscillators are initially in a thermal state, see Appendix~\ref{app_groundstate} for details. Conversely, depending on the sensing case, the spin state will be initialized differently. Note that at $\tau$ multiples of $2\pi$, the unitary operator $\hat{U}(\tau{=}2\pi)$ acquires relative phases for the spin states ${\propto}e^{2\pi i \hat{Z}(\bm{k}{,}g)^2}$, whereas the coherent field returns to its initial state $|\alpha\rangle$. For $0{<}\tau{<}2\pi$, the mechanical state rotates in phase-space by an angle $\tau$ due to $e^{-i\hat{a}^\dagger\hat{a}\tau}$, and it is conditionally displaced by the spin state by an amount ${\propto}\langle \hat{Z}(\bm{k},g)\rangle\eta(\tau)$. Here, $\langle\cdot\rangle$ is the expected value with respect to the spin state. In what follows, we will investigate the precision limits of gravimetry for equal-scaled spin-mechanical coupling $k_i{=}k_j, \forall i {,}j$. The scenario where $k_i{\neq}k_j, \forall i{,}j$ will be addressed later.

\section{Elemental gravimetry scenarios}\label{sec_elementalgravimetry}

Two straightforward scenarios can be explored: (i) a single spin and (ii) two spins coupled to the mechanical oscillator. The former, initialized as $|\psi(0)\rangle{=}\frac{1}{\sqrt{2}}(|0\rangle{+}|1\rangle)|\alpha\rangle$, yields a QFI (see Appendix~\ref{app_QFIsm_N1_spins} for details):
\begin{equation}
  \mathcal{Q}_\mathrm{sm}^{(1)}\propto k^2 (\tau - \sin\tau)^2 - 2\cos \tau.\label{eq_qsm1}
\end{equation}
For the two spins scenario, we initialized the spins in their computational basis: $|\psi(0)\rangle_\mathrm{spin}^{(2)}{=}R_1|01\rangle{+}R_2|00\rangle{+}R_3|10\rangle{+}R_4|11\rangle$, with $R_j{=}r_je^{i\phi_j}$ and $\sum\limits_{j{=}1}^4|R_j|^2{=}1$. Giving the QFI (see Appendix~\ref{app_QFIsm_N2_spins} for details):
\begin{equation}
\mathcal{Q}_\mathrm{sm}^{(2)}{\propto}8 k^2[r_4^2{-}r_4^4{-}r_2^4{+}r_2^2 (1{+}2 r_4^2)](\tau{-}\sin\tau)^2{-}4\cos\tau.\label{eq_qsm2}
\end{equation}
Maximizing Eq.~\eqref{eq_qsm2} over the values of $r_2$ and $r_4$ yields $r_2{=}r_4{=}1/\sqrt{2}$. Hence, the triplet state $\sqrt{2}|\psi(0)\rangle_\text{spin}^{(2)}{=}(|00\rangle{+}|11\rangle)$~\cite{chen2018highprecision} gives rise to $\mathcal{Q}_\text{sm}^{(2)}{\propto}4k^2(\tau{-}\sin\tau)^2{-}2\cos\tau$, which can be interpreted as the single spin scenario with an effective coupling of $2k$. As experimental constraints limit the single spin-mechanical coupling strength $k$, coupling more spins to the mechanical object might alleviate this restriction. The enhancement in gravimetry precision provided by the triplet state motivates us to explore the ultimate limits of gravimetry precision with respect to $N$ spins.

\section{Heisenberg-limited gravimetry}\label{sec_SM_heisenberg}

\begin{figure}[t]
    \centering
    \includegraphics[width=\linewidth]{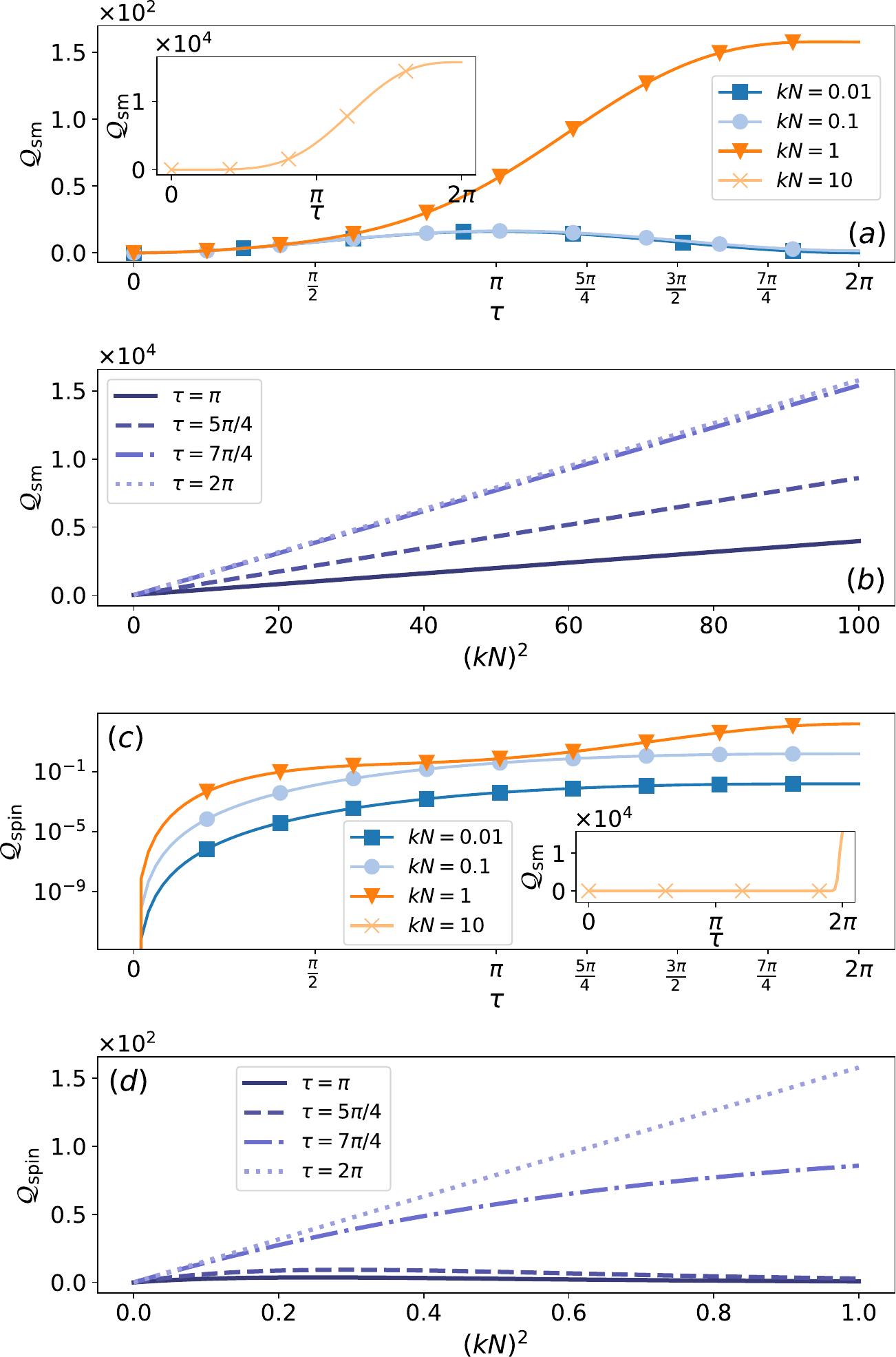}
    \caption{(a) Quantum Fisher information (QFI) of the spin-mechanical system $\mathcal{Q}_\mathrm{sm}$ as a function of time $\tau$ for several values of $kN$. (b) $\mathcal{Q}_\mathrm{sm}$ as a function of $(kN)^2$ for several times $\tau$. (c) QFI of the spin system $\mathcal{Q}_\mathrm{spin}$ as a function of time $\tau$ for several values of $kN$. (d) $\mathcal{Q}_\mathrm{spin}$ as a function of $(kN)^2$ for several times $\tau$.}
    \label{fig_GHZ_limits}
\end{figure}

We introduce the collective spin operator $\hat{S}_\beta{=}\sum\limits_{i{=}1}^N\frac{\hat{\sigma}_i^\beta}{2}$. Fulfilling: $[\hat{S}_n{,}\hat{S}_{m}]{=}i\hbar\varepsilon_{nml}\hat{S}_l$, $\hat{S}^2|s{,}m\rangle{=}\hbar^2s(s{+}1)|s{,}m\rangle$ and $\hat{S}_z|s{,}m\rangle{=}\hbar m|s{,}m\rangle$. In the above, $\varepsilon_{nml}$ is the Levi-Civita symbol, $s{\leq}N{/}2$ is the spin number, and ${-}s{\leq}m{\leq}s$ is the angular momentum's component along the $z$-direction ($|m|{\leq}s$). With this notation, $\hat{Z}(\bm{k}{,}g){=}k\hat{S}_z{-}g$. By considering the symmetric case, namely $2s{=}N$, and an initial spin-mechanical state $\ket{\psi(0)}{=}\sum\limits_{m{=}-N/2}^{N/2}c_m|\frac{N}{2}{,}m\rangle\ket{\alpha}$ with $c_m$ a complex coefficient, we obtain the spin-mechanical evolved state:
\begin{equation}
    \ket{\psi(\tau)}_\mathrm{sm}{=}\sum\limits_{m{=}-\frac{N}{2}}^{\frac{N}{2}}\tilde{C}_m\ket{\frac{N}{2}{,}m}\ket{\varphi(m{,}g)},\label{eq_psi_m}
\end{equation}
where $\tilde{C}_m{:=}c_m e^{imk[(\tau{-}\sin\tau)(km{-}2g){+}\alpha\sin\tau]}$, and the mechanical amplitude $\varphi(m{,}g){:=}\alpha e^{-i\tau}{+}(km{-}g)\eta(\tau)$ is conditionally displaced by a quantity $(km{-}g)\eta(\tau)$. The wave function of Eq.~\eqref{eq_psi_m} is general. In particular, for the Greenberger-Horne-Zeilinger (GHZ) state ($m{=}\pm\frac{N}{2}$)~\cite{shammah2018open}:
\begin{equation}
\ket{\psi(0)}_\mathrm{spin}^{(\text{GHZ})}{=}\frac{1}{\sqrt{2}}\left(\ket{\frac{N}{2}{,}\frac{N}{2}}{+}\ket{\frac{N}{2}{,}-\frac{N}{2}}\right),
\end{equation}
the QFI is found to be, see Appendix~\ref{app_QFIsm_GHZ} for details:
\begin{equation}
\mathcal{Q}_\mathrm{sm}^{(\text{GHZ})}{\propto}k^2 N^2 (\tau{-}\sin\tau)^2{-}2 \cos\tau.\label{eq_GHZ_Heisenberg_Limit_sm}
\end{equation}
Eq.~\eqref{eq_GHZ_Heisenberg_Limit_sm} demonstrates that the spin-mechanical probe scales quadratically with the number of particles $N$ for any given time, i.e., it reaches the Heisenberg limit for gravimetry precision. This is the first main result of this work. Note that this result is non-trivial. In phase-like sensing tasks with Hamiltonian $\hat{H}{=}\Xi\hat{h}_0$, where $\hat{h}_0$ is a Hermitian generator~\cite{paris2009quantum} and $\Xi$ is a unknown parameter to be estimated, the GHZ state has been established as the optimal choice for achieving the Heisenberg limit~\cite{giovannetti2004quantum}. However, for quantum probes governed by general Hamiltonian $\hat{H}{\neq}\Xi\hat{h}_0$, the optimal state will depend on the sensing task~\cite{zwierz2010general, fiderer2019maximal}. Interestingly, Eq.~\eqref{eq_GHZ_Heisenberg_Limit_sm} resembles the single spin scenario with an amplified effective spin-mechanical coupling strength of $kN$. To demonstrate this, in Fig.~\ref{fig_GHZ_limits}(a), we plot the QFI of the spin-mechanical system $\mathcal{Q}_\mathrm{sm}$ as a function of time $\tau$ for several values of $kN$. As the figure shows, the $\mathcal{Q}_\mathrm{sm}$ benefits from increasing $kN$. In Fig.~\ref{fig_GHZ_limits}(b), we plot the $\mathcal{Q}_\mathrm{sm}$ as a function of $(kN)^2$ for several times $\tau$. As shown in the figure, $\mathcal{Q}_\mathrm{sm}$ achieves the Heisenberg limit of precision, that is increases quadratically with respect to the number of particles $N$ for any given time. For values of $kN{\lesssim}0.1$, the $\mathcal{Q}_\mathrm{sm}$ peaks around $\tau{\sim}\pi$, where the spins are highly entangled with the field~\cite{montenegro2014nonlinearity}. Thus, the gravitational information $g$ is encoded in the entire correlated state. On the other hand, for $kN{\gtrsim}1$, $\mathcal{Q}_\mathrm{sm}$ peaks at $\tau{\rightarrow}2\pi$, where the field completely disentangles from the spins. Consequently, all information regarding gravitational acceleration $g$ is encoded in the spins subsystem. 

The above entanglement facts can be observed from the linear entropy \begin{equation}
    S_L = 1 - \text{Tr}(\rho_m^2).~\label{eq_linear_entropy}
\end{equation}
Here, $\rho_m$ is the reduced density matrix for the mechanical object, namely $\rho_m = \text{Tr}_s(\rho_\mathrm{sm})$---$\text{Tr}_s$ is the partial trace with respect to the spin subsystem. It is then straightforward to evaluate $S_L$ from Eq.~\eqref{eq_linear_entropy} and Eq.~\eqref{eq_psi_m} as:
\begin{equation}
S_L{=}\left(1{-}e^{2k^2N^2(\cos\tau{-}1)}\right)/2.
\end{equation}

\begin{figure}[t]
    \centering
    \includegraphics[width=\linewidth]{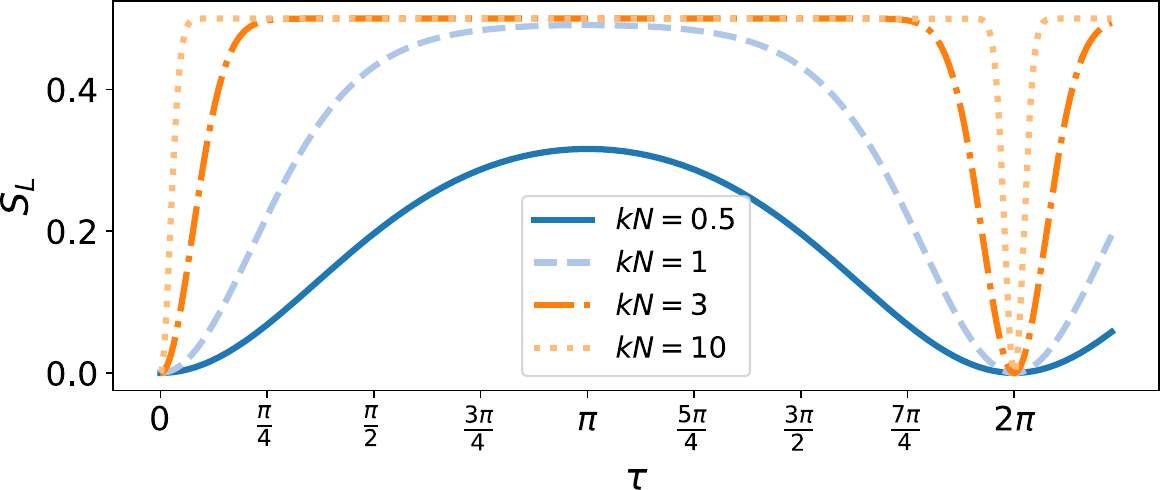}
    \caption{Linear entropy $S_L$ as a function of time $\tau$ for various $kN$ values. Entanglement increases with $kN$ and saturates for $kN \gtrsim 1$, with the system disentangling at $\tau = 2\pi$ multiples.}
    \label{fig_sm_Entropy}
\end{figure}

\noindent Although entangled states are typically invoked to achieve quantum-enhanced sensitivity~\cite{degen2017quantum, montenegro2024reviewquantummetrologysensing}, in this case, it is the spin-mechanical disentanglement that transfers all the information about $g$ to the spin state. This is the first indication that the Heisenberg limit can indeed be achieved by partially probing only the spin subsystem. In Fig.~\ref{fig_sm_Entropy}, we plot the linear entropy $S_L$ as a function of time $\tau$ for several values of $kN$. As the figure shows, increasing the value of $kN$ leads to increased entanglement, which saturates for values $kN{\gtrsim}1$. Interestingly, regardless of the values of $kN$, the system disentangles at multiples of $\tau = 2\pi$. Note that the disentanglement point provides the highest value of QFI, as opposed to the highly entangled state typically used for quantum enhancement. This is because the inherent parametric nature of the spin-mechanical system dynamically transfers all the information encoded in the field mode to the spin subsystem.

\section{Heisenberg-limited spin gravimeter}\label{sec_S_heisenberg}

Probing the entire system can be practically challenging~\cite{montenegro2022probing}. Moreover, the maximum QFI occurs when the spin disentangles from the field, suggesting that the Heisenberg limit can be achieved by partially probing only the spins. Therefore, we focus on the gravimetry precision for the spins subsystem. The QFI of the spin subsystem, see Appendix~\ref{app_QFIs_Nspins} for details, is:
\begin{equation}
    \mathcal{Q}_\mathrm{spin}{=}4e^{2 k^2N^2(\cos\tau{-}1)}k^2N^2(\tau{-}\sin\tau)^2.\label{eq_Q_spin}
\end{equation}
Note that at $\tau$ multiples of $2\pi$, $\mathcal{Q}_\mathrm{spins}{\propto}k^2N^2\tau^2$ and the spin subsystem reaches the Heisenberg limit of precision, acquiring all the information regarding $g$, while the mechanical oscillator returns to its initial state $\ket{\alpha}$. To illustrate this in detail, in Fig.~\ref{fig_GHZ_limits}(c), we plot $\mathcal{Q}_\mathrm{spin}$ as a function of time $\tau$ for different values of $kN$. The figure reveals two key features: (i) $\mathcal{Q}_\mathrm{spin}$ always peaks at $\tau$ multiples of $2\pi$, and (ii) as $kN$ increases, $\mathcal{Q}_\mathrm{spin}$ becomes vanishingly small, except at $\tau{\rightarrow}2\pi$ [see inset of Fig.~\ref{fig_GHZ_limits}(c)]. Indeed, in Fig.~\ref{fig_GHZ_limits}(d), we plot $\mathcal{Q}_\mathrm{spin}$ as a function of $(kN)^2$ for various choices of times $\tau$. As depicted in the figure, only at $\tau{=}2\pi$ the $\mathcal{Q}_\mathrm{spin}$ exhibits the Heisenberg limit. For other values of $\tau$, $\mathcal{Q}_\mathrm{spins}$ decreases exponentially by the positive function $e^{2 k^2N^2(\cos\tau{-}1)}$ as $kN$ increases, thereby losing the Heisenberg limit of precision. 

Two important points need clarification: (i) the failure to tune (or prepare) the probe~\cite{kacprowicz2010experimental} to the parameter that achieves the Heisenberg limit of precision, here $\tau{=}2\pi$, is a common challenge in \textit{local} quantum sensing~\cite{rams2018atthelimits, montenegro2021global, chiranjibglobal}; and (ii) losing the Heisenberg limit of precision should not be confused with losing quantum advantage entirely. To illustrate this, we examine the gravimetry precision limits for both a classical and a quantum probe in the vicinity of $\tau{=}2\pi$, where the quantum probe outperforms the classical probe in most cases, see later Section \textit{``Robustness in the vicinity of $\tau{=}2\pi$"} for details.

\section{Classical Fisher information}\label{sec_CFI}

\begin{figure}[t]
    \centering
    \includegraphics[width=\linewidth]{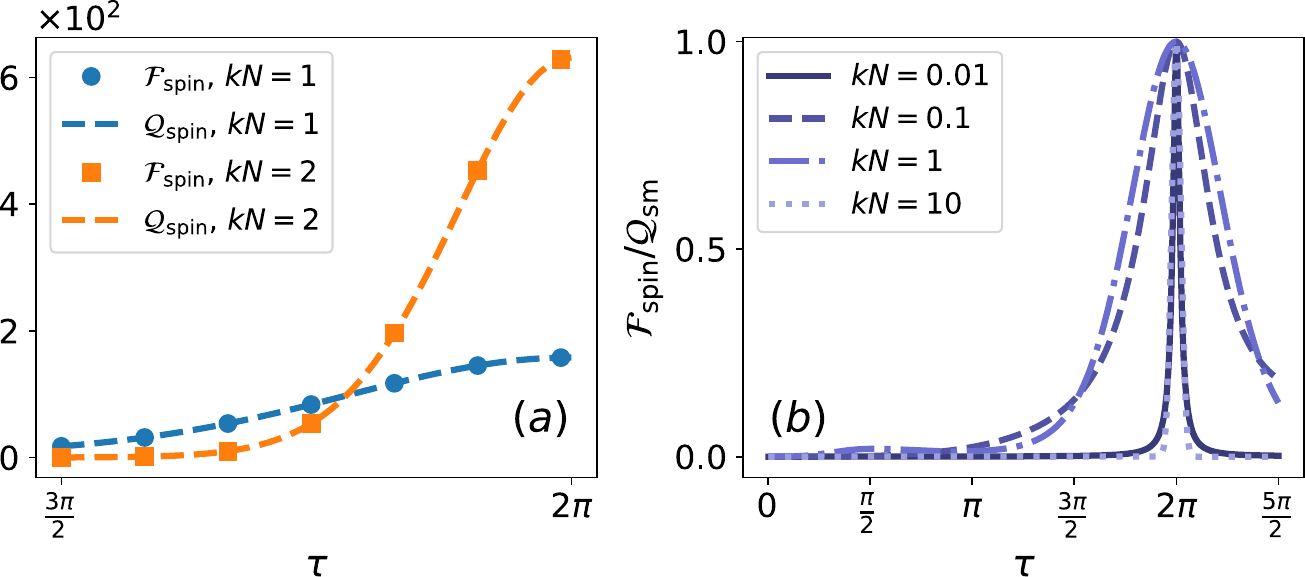}
    \caption{(a) Spin QFI $\mathcal{Q}_\mathrm{spin}$ and the spin CFI $\mathcal{F}_\mathrm{spin}$ as a function of time $\tau$ for several choices of $kN$. (b) Ratio $\mathcal{F}_\mathrm{spin}{/}\mathcal{Q}_\mathrm{sm}$ as a function of time $\tau$ for several choices of $kN$.}
    \label{fig_QFIs_ratios}
\end{figure}

\begin{figure}[t]
    \centering
    \includegraphics[width=\linewidth]{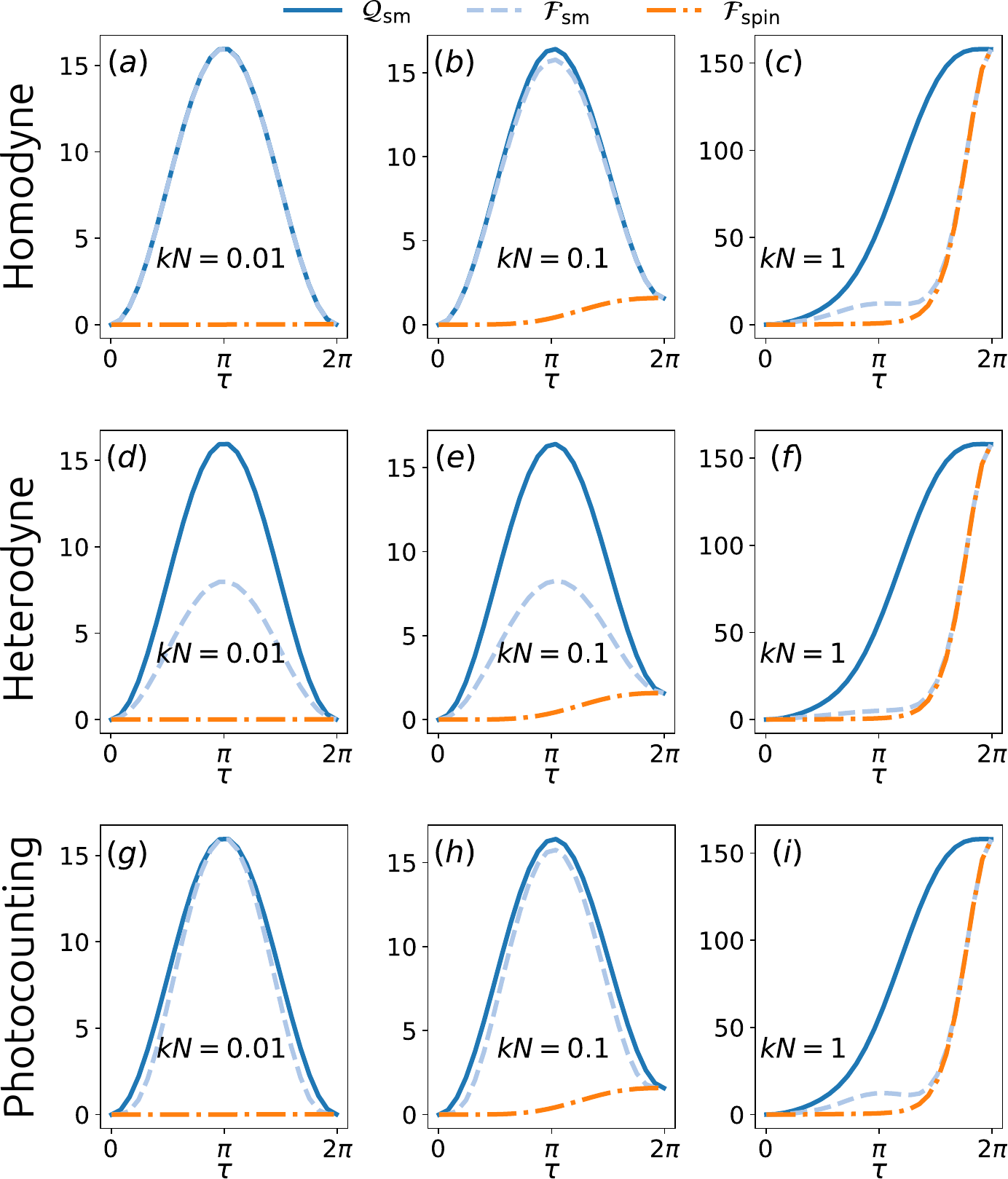}
    \caption{All panels depict the CFI obtained solely from measurements on the spin subsystem $\mathcal{F}_\mathrm{spin}$ and the ultimate gravimetry precision quantified by the spin-mechanical QFI $\mathcal{Q}_\mathrm{sm}$ as a function of time $\tau$ for different values of $kN$. Panels (a)-(c) include the CFI from uncorrelated measurements on the spin and the field via homodyne detection $\mathcal{F}_\mathrm{sm}^{\mathrm{hom.}}$; panels (d)-(f) consider the CFI obtained from uncorrelated heterodyne measurements and spin measurements $\mathcal{F}_\mathrm{sm}^{\mathrm{het.}}$; and panels (g)-(i) show the CFI obtained from uncorrelated photocounting measurements and spin measurements $\mathcal{F}_\mathrm{sm}^{\mathrm{pho.}}$.}
    \label{fig_sm_CFIs}
\end{figure}

Obtaining the optimal measurement to reach the QFI of the entire system, as derived from the SLD $L_g$, is often difficult or even impractical~\cite{paris2009quantum}. Thus, finding a way to measure a subsystem that still captures the maximum possible information about the unknown parameter is essential. Here, we show that by dynamically disentangling the spin from the mechanics, a local, feasible measurement can achieve the QFI of the entire system, i.e., the ultimate precision limit of the probe. To this end, we construct the probability distributions required to evaluate the CFI for the spin subsystem using a POVM given by: $\hat{\Pi}_\Upsilon{=}|\Upsilon(\Theta,\Phi)\rangle\langle\Upsilon(\Theta,\Phi)|$, where $|\Upsilon(\Theta,\Phi)\rangle{=}\cos\frac{\Theta}{2}\ket{\frac{N}{2},\frac{N}{2}}{+}\sin\frac{\Theta}{2}e^{-i\Phi}\ket{\frac{N}{2}{,}{-}\frac{N}{2}}$. With this choice of spin POVM, $\Upsilon$ can take two measurement outcomes, and the optimal CFI is (see Appendix~\ref{app_CFIs_analytical} for its analytical expression):
\begin{equation}
    \mathcal{F}_\mathrm{spin}{=}\max_{ \{\Theta,\Phi\}} \left[\frac{\left[\partial_g p(\Upsilon(\Theta{,}\Phi)|g)\right]^2}{p(\Upsilon(\Theta{,}\Phi)|g)[1{-}p(\Upsilon(\Theta{,}\Phi)|g)]}\right],
\end{equation}
where $p(\Upsilon(\Theta{,}\Phi)|g){=}\bra{\psi(\tau)}\hat{\Pi}_\Upsilon\ket{\psi(\tau)}$ is the probability associated to POVM $\hat{\Pi}_\Upsilon$. Maximization analysis shows that $\Theta{=}\frac{\pi}{2}$, while $\Phi$ takes other values. Note that projective measurements of the type $\{\hat{\Pi}_\Upsilon\}$ have been proposed for verifying GHZ states~\cite{zhao2021creation}, and the evaluation of the magnetization function through measurements of individual particles has been successfully performed in experiments~\cite{wang2009individual, monroe2021programmable, islam2011onset, islam2013emergence, richerme2013quantum, lee2016engineering, mooney2021generation}. In Fig.~\ref{fig_QFIs_ratios}(a), we compare the spin QFI $\mathcal{Q}_\mathrm{spin}$ and the spin CFI $\mathcal{F}_\mathrm{spin}$ as a function of time $\tau$ for several choices of $kN$. As the figure shows, the choice of the POVM $\hat{\Pi}_\Upsilon$ optimized over the angles $\{\Theta{,}\Phi\}$ saturates the QFI of the spin subsystem. Remarkably, for the specific case of $\tau{=}2\pi$ (the spin-mechanical disentanglement time), $\hat{\Pi}_\Upsilon$ also saturates the QFI of the entire spin-mechanical system. Hence, it constitutes the optimal measurement basis for the entire spin-mechanical system at $\tau{=}2\pi$, achieving the Heisenberg gravimetry precision limit. Achieving the Heisenberg limit with a local and feasible spin measurement is the second key result of this work. To quantify the sensing performance of probing the spin subsystem relative to the entire spin-mechanical system for times $\tau{\neq}2\pi$, in Fig.~\ref{fig_QFIs_ratios}(b), we plot the ratio $\mathcal{F}_\mathrm{spin}{/}\mathcal{Q}_\mathrm{sm}{\leq}1$ as a function of time $\tau$ for several choices of $kN$. The figure reveals two key observations: (i) The information fraction from the spin subsystem is notably low for times $0{<}\tau{\lesssim}2\pi$ in the extreme cases of $kN{\ll}1$ and $kN{\gg}1$ [in agreement with Eq.~\eqref{eq_Q_spin}], and (ii) all information about $g$ becomes encoded in the spin subsystem as the field dynamically disentangles from the spin, with $\mathcal{F}_\mathrm{spin}{=}\mathcal{Q}_\mathrm{sm}$ for the separable spin-mechanical state at $\tau{=}2\pi$. This suggests that there is an optimal value of $kN$ that maximizes the ratio $\mathcal{F}_\mathrm{spin}{/}\mathcal{Q}_\mathrm{sm}$ in the vicinity of $\tau{=}2\pi$. 
Here, we assume access only to the spin subsystem. However, additional information about the unknown parameter $g$ could be gained by extra measurements on the mechanical mode (For the sake of an experimentally friendly approach, we exclude correlated spin-mechanical measurements; recall that the optimal measurements are derived from the eigenstates of the SLDs $L_g$). When access to the mechanical mode is possible, measurements are typically restricted to homodyne, heterodyne, or photocounting techniques. For the homodyne approach, the joint CFI is:
\begin{equation} 
    \mathcal{F}_\mathrm{sm}^{\mathrm{hom.}}{=}\max_{ \{\Theta,\Phi\}} \left[\int_{-\infty}^\infty\frac{\left(\partial_g p(\Upsilon(\Theta{,}\Phi), x|g)\right]^2}{p(\Upsilon(\Theta{,}\Phi), x|g)[1{-}p(\Upsilon(\Theta{,}\Phi), x|g)]}dx\right],
\end{equation}
where $p(\Upsilon(\Theta{,}\Phi), x|g){=}\bra{\psi(\tau)}(\hat{\Pi}_\Upsilon{\otimes}\ket{x_\lambda}\langle x_\lambda|)\ket{\psi(\tau)}$ is the conditional probability associated to POVMs $\hat{\Pi}_\Upsilon{\otimes}\ket{x_\lambda}\langle x_\lambda|$ and $\ket{x_\lambda}$ is the eigenstate of the rotated quadrature $\sqrt{2}\hat{x}_\lambda{=}\hat{a}e^{-i\lambda}{+}\hat{a}^\dagger e^{i\lambda}$. In Fock basis representation it reads $\langle n|x_\lambda\rangle{=}\pi^{-1/4}2^{-n/2}(n!)^{-1/2}\mathrm{exp}[-x_\lambda^2/2]\mathcal{H}_n(x_\lambda)\mathrm{exp}[in\lambda]$; where $n$ is the Fock number state and $\mathcal{H}_n(x_\lambda)$ are the Hermite polynomials of order $n$. In Fig.~\ref{fig_sm_CFIs}, we compare the CFI from both measuring the spin subsystem ($\mathcal{F}_\mathrm{spin}$) and the spin-mechanical system ($\mathcal{F}_\mathrm{sm}^{\mathrm{hom.}}$) with the ultimate gravimetry precision given by the QFI of the entire system ($\mathcal{Q}_\mathrm{sm}$) as a function of time $\tau$ and different values of $kN$. As seen from Figs.~\ref{fig_sm_CFIs}(a){-}(b), for $kN{\lesssim}0.1$, most of the information content with respect to $g$ is encoded in the mechanical subsystem, which can be fairly extracted using homodyne detection. In Fig.~\ref{fig_sm_CFIs}(c), for values $kN{\gtrsim}1$, measuring the field via homodyne detection results in poor performance across all times. Measuring only the spin subsystem at and around $\tau{\rightarrow}2\pi$ yields excellent performance.
For heterodyne detection, the conditional probability needed to evaluate the CFI is:
\begin{equation}
    p(\Upsilon(\Theta{,}\Phi), \zeta|g)=\frac{1}{\pi}\bra{\psi(\tau)} (\hat{\Pi}_\Upsilon\otimes\ket{\zeta}\bra{\zeta}) | \psi(\tau) \rangle,
\end{equation}
where $\ket{\zeta}$ is a coherent state with $\zeta \in \mathbb{C}$. Therefore, the CFI is computed as:
\begin{equation}
    \mathcal{F}_\mathrm{spin}^{\mathrm{het.}}=\max_{ \{\Theta,\Phi\} }\left[\int d^2\zeta \frac{\left(\partial_g p(\Upsilon(\Theta{,}\Phi), \zeta|g)\right)^2}{p(\Upsilon(\Theta{,}\Phi), \zeta|g)[1{-}p(\Upsilon(\Theta{,}\Phi), \zeta|g)]}\right].
\end{equation}
Similarly, for the photocounting scheme, the conditional probability is:
\begin{equation}
    p(\Upsilon(\Theta{,}\Phi), n|g)=\bra{\psi(\tau)} (\hat{\Pi}_\Upsilon\otimes\ket{n}\bra{n}) | \psi(\tau) \rangle,
\end{equation}
where $\ket{n}$ is a Fock number state. Hence:
\begin{equation}
    \mathcal{F}_\mathrm{spin}^{\mathrm{pho.}}=\max_{ \{\Theta,\Phi\} }\left[\sum_{n=0}^\infty \frac{\left(\partial_g p(\Upsilon(\Theta{,}\Phi), n|g)\right)^2}{p(\Upsilon(\Theta{,}\Phi), n|g)[1{-}p(\Upsilon(\Theta{,}\Phi), n|g)]}\right].
\end{equation}

In Figs.~\ref{fig_sm_CFIs}(d)-(f), we plot the CFI obtained from uncorrelated heterodyne measurements and spin measurements $\mathcal{F}_\mathrm{sm}^{\mathrm{het.}}$ as a function of time $\tau$ for several values of $kN$. In Figs.~\ref{fig_sm_CFIs}(g)-(i), we plot the CFI obtained from uncorrelated photocounting measurements and spin measurements $\mathcal{F}_\mathrm{sm}^{\mathrm{pho.}}$ as a function of time $\tau$ and different $kN$. The figures show that the homodyne measurement scheme generally outperforms the other mechanical probing methods.

\section{Robustness against anisotropies}\label{sec_robustness_anisotropies}

\begin{figure}[t]
    \centering
    \includegraphics[width=\linewidth]{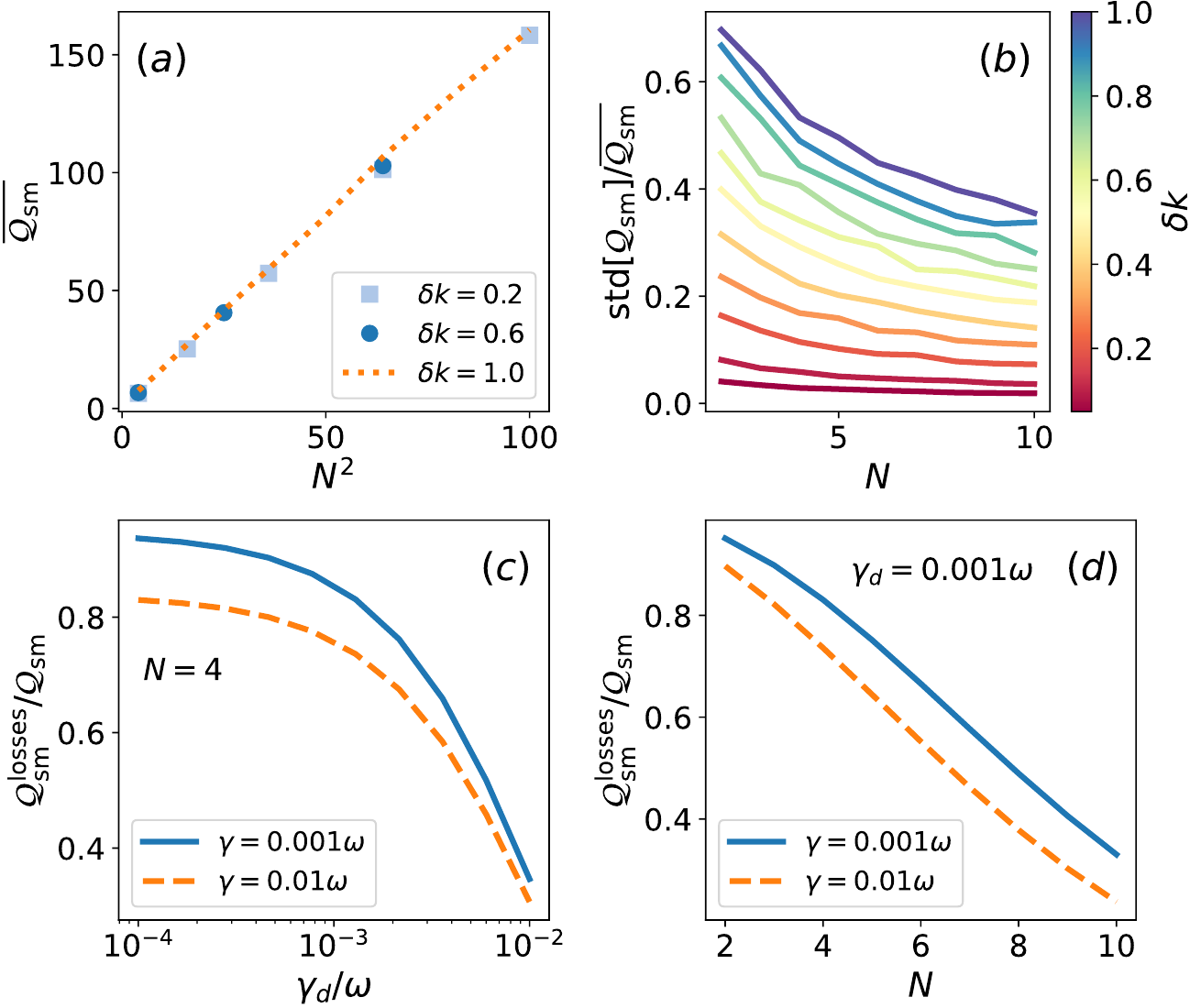}
    \caption{(a) Average QFI $\overline{\mathcal{Q}_\mathrm{sm}}$ as a function of $N^2$ for several choices of $\delta k$. (b) Standard deviation $\mathrm{std}[\mathcal{Q}_\mathrm{sm}]$ divided by $\overline{\mathcal{Q}_\mathrm{sm}}$ as a function of $N$ for several choices of $\delta k$. (c) Comparison at $\tau{=}2\pi$ between the lossless spin-mechanical QFI $\mathcal{Q}_\mathrm{sm}$ with the losses spin-mechanical QFI $\mathcal{Q}_\mathrm{sm}^{\mathrm{losses}}$ as a function of the collective dephasing rate $\gamma_d$ for different collective emission rates $\gamma$ and $N{=}4$. (d) Fraction $\mathcal{Q}_\mathrm{sm}^{\mathrm{losses}}/\mathcal{Q}_\mathrm{sm}$ at $\tau{=}2\pi$ as a function of $N$ for two values of collective emission $\gamma$ and fixed collective dephasing $\gamma_d{=}10^{-3}\omega$. Panels (c) and (d) consider $k{=}1.$}
    \label{fig_robustness}
\end{figure}

We now investigate the quantum probe robustness in the presence of unequal spin-mechanical coupling strengths $k_i{\neq}k_j, \forall i{,}j$. We simulate the dynamics under $\hat{Z}(\bm{k}{,}g){:=}\left[\sum\limits_{i=1}^N k_i \frac{\sigma_i^z}{2}{-}g\right]$ starting from $\sqrt{2}\ket{\psi(0)}_\mathrm{sm}{=}(\ket{0}^{\otimes N}{+}\ket{1}^{\otimes N})\ket{\alpha}$. Each coupling strength fluctuates randomly around a central value $k_i$, with $k_i{\rightarrow}k_i(1{+}\Delta k_i)$, where $\Delta k_i$ is a random number in the interval $[-\delta k{,}\delta k]$. In Fig.~\ref{fig_robustness}(a), we plot the average QFI over 1000 instances $\overline{\mathcal{Q}_\mathrm{sm}}$ as a function of $N^2$ for several random amplitudes $\delta k$. As seen in the figure, the Heisenberg gravimetry precision holds on average. However, in Fig.~\ref{fig_robustness}(b), we plot the ratio between the standard deviation $\mathrm{std}[\mathcal{Q}_\mathrm{sm}]$ (from sampling 1000 instances) and the average $\overline{\mathcal{Q}_\mathrm{sm}}$ as a function of $N$ for several choices of $\delta k$. Ideally, one wants $\mathrm{std}[\mathcal{Q}_\mathrm{sm}]{\ll}\overline{\mathcal{Q}_\mathrm{sm}}$. The figure shows strong statistical deviations from $\overline{\mathcal{Q}_\mathrm{sm}}$ as $\delta k$ increases. Remarkably, these deviations decrease as $N$ increases for a fixed $\delta k$. Thus, with unequal random spin-mechanical coupling strengths, gravimetry precision maintains the Heisenberg limit but faces increased uncertainty in standard deviation as the random amplitude $\delta k$ grows.

\section{Decoherence analysis}\label{sec_decoherence}
Although a full analysis of the open quantum system is beyond our scope, we present a simple examination of the decoherence channels involved in the spin-mechanical dynamics. To achieve this, we solve the Born-Markov master equation in Lindblad form as follows:
\begin{multline}
\frac{d\rho(\tau)}{d\tau}{=}{-}i[\hat{H}{,}\rho(\tau)]{+}\gamma_d \left(\hat{S}_z\rho(\tau)\hat{S}_z^\dagger{-}\frac{1}{2}\{ \hat{S}_z^\dagger\hat{S}_z{,}\rho(\tau)\}\right)\\
{+}\gamma\left(\hat{S}_-\rho(\tau)\hat{S}_+{-}\frac{1}{2}\{\hat{S}_+\hat{S}_-{,}\rho(\tau)\}\right){+}\kappa n_{\text{th}}\Big(\hat{a}^\dagger\rho(\tau)\hat{a}\\
{-}\frac{1}{2}\{\hat{a}\hat{a}^\dagger{,}\rho(\tau)\}\Big){+}\kappa(n_{\text{th}}{+}1)\left(\hat{a}\rho(\tau)\hat{a}^\dagger{-}\frac{1}{2}\{\hat{a}^\dagger\hat{a}{,}\rho(\tau)\}\right),\label{eq_master_equation}
\end{multline}
where we consider a reservoir with an average of $n_\mathrm{th}$ excitations, along with a collective dephasing, collective emission, and mechanical damping with decay rates $\gamma_d$, $\gamma$, and $\kappa$, respectively. Due to the excellent mechanical quality factors~\cite{aspelmeyer2014cavity, treutlein2014hybrid}, we consider a modest value of $\kappa{=}10^{-5}\omega$, and a reservoir with an average of $n_\mathrm{th}{=}10$ excitations. In Fig.~\ref{fig_robustness}(c), we compare the lossless spin-mechanical QFI $\mathcal{Q}_\mathrm{sm}$ from Eq.~\eqref{eq_GHZ_Heisenberg_Limit_sm} with the losses spin-mechanical QFI $\mathcal{Q}_\mathrm{sm}^{\mathrm{losses}}$ (obtained from the state by solving Eq.~\eqref{eq_master_equation}) as a function of the collective dephasing rate $\gamma_d$ for different collective emission rates $\gamma$ and for a given number of particles $N{=}4$. As seen in the figure, a fair fraction ${\sim}0.9$ of the ultimate sensing precision at $\tau{=}2\pi$ can be achieved for values of $\gamma_d{=}\gamma{=}10^{-3}\omega$. To study how the fraction $\mathcal{Q}_\mathrm{sm}^{\mathrm{losses}}/\mathcal{Q}_\mathrm{sm}$ behaves with the number of spin particles $N$, we plot this fraction as a function of $N$ in Fig.~\eqref{fig_robustness}(d) for two values of collective emission $\gamma$ and a fixed collective dephasing $\gamma_d{=}10^{-3}\omega$. As the figure shows, as $N$ increases, the fraction decreases, signaling a negative impact for gravimetry. Note that several other dissipation channels, particularly local decays~\cite{shammah2018open}, must be considered. Despite the negative results above, this analysis helps identify the decoherence values at which the proposed gravimetry sensing scheme remains feasible. Techniques for preparing deterministic states dissipatively~\cite{poyatos1996quantum, kastoryano2011dissipative, li2018dissipation, cho2011optical, stannigel2012driven, de2017steady, groszkowski2022reservoir} and metrology in noisy scenarios~\cite{escher2011general, falaye2017investigating, albarelli2018restoringheisenberg} could help improve this technical issues.

\section{Robustness in the vicinity of $\tau = 2\pi$}\label{sec_robustness_vicinity}

\begin{figure}[t]
\centering\includegraphics[width=\linewidth]{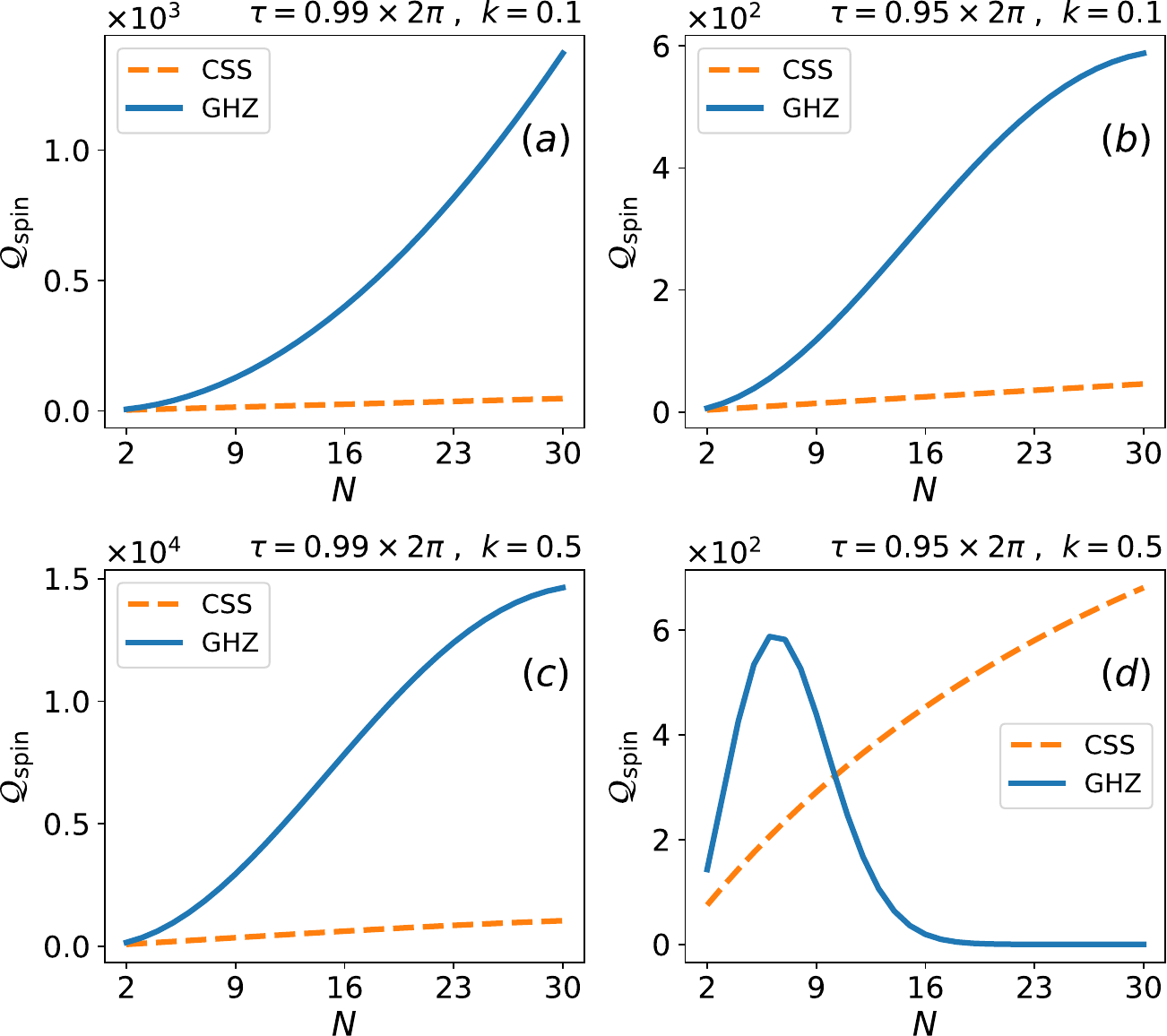}
    \caption{QFI for the reduced density matrix of the spin subsystem, Eq.~\eqref{eq_Q_spin}. (Blue solid curve) The system evolves from the initial state $|\psi(0)\rangle{=}|\mathrm{GHZ}\rangle|\alpha\rangle$; (Orange dashed curve) The system evolves from the initial state $ |\psi(0)\rangle{=}|\mathrm{CSS}\rangle|\alpha\rangle $. We consider $\alpha{=}0$, two values for $k{=}0.1$ and $k{=}0.5$, and two values in the vicinity of $\tau{=}2\pi$, specifically $\tau{=}0.99{\times}2\pi$ and $\tau{=}0.95{\times}2\pi$. Even in the absence of the Heisenberg limit, the $\mathrm{GHZ}$ scenario significantly outperforms the $\mathrm{CSS}$ scenario in several instances.}
    \label{fig_sm_vicinity}
\end{figure}

Achieving the Heisenberg limit of precision faces two common challenges: (i) extreme sensitivity to particle loss, e.g., losing even one particle in GHZ/N00N-type states, $|\text{N00N}\rangle{\sim}|\text{N,0}\rangle{+} |\text{0,N}\rangle$, can destroy the maximum sensing precision~\cite{kacprowicz2010experimental}; and (ii) the need for precise tuning of the parameters that allow this limit to be reached~\cite{rams2018atthelimits}. Since the Heisenberg limit is reached at multiples of $\tau{=}2\pi$ by locally probing the spin subsystem, it is pertinent to investigate the robustness of this precision limit near $\tau{=}2\pi$, especially given the exponentially decreasing factor $e^{2 k^2 N^2 (\cos \tau{-}1)}$ in Eq.~\eqref{eq_Q_spin}. In Figs.~\ref{fig_sm_vicinity}, we evaluate Eq.~\eqref{eq_Q_spin} as a function of $N$ for different values of $k$ and times in the vicinity of $\tau{=}2\pi$. As shown in the figures, the Heisenberg limit is rapidly lost near $\tau{=}2\pi$. Although a super-linear scaling is observed in Fig.~\ref{fig_sm_vicinity}(a), this behavior quickly diminishes when either the coupling constant $k$ increases or the time $\tau$ deviates from $2\pi$. This is clearly illustrated in Fig.~\ref{fig_sm_vicinity}(d), where for $k{=}0.5$ and $\tau{=}0.95 \times 2\pi$, the term $e^{2 k^2 N^2 (\cos \tau{-}1)}$ in Eq.~\eqref{eq_Q_spin} becomes dominant. Hence, for $N{>}10$, the curve vanishes very quickly.

Nonetheless, it is important not to confuse losing the Heisenberg limit of precision with losing any quantum enhancement in precision. Even if the Heisenberg limit is not achieved, multiplicative factors $a$ in the scaling of the QFI, $\mathcal{Q}{\sim}a N^b$, can still result in higher values compared to classical probes of which $\mathcal{Q}{\sim}c N$, where $c$ is a real coefficient. Consequently, the quantum probe can still offer better precision than classical probes. To address this issue, we evaluate the QFI from a purely classical spin state, defined as follows~\cite{pezze2018quantum}:
\begin{multline}
    |\mathrm{CSS}\rangle{=}\bigotimes_{j=1}^{N}\frac{1}{\sqrt{2}}(|0\rangle{+}|1\rangle)_j\\
    {=}\sum_{m=-\frac{N}{2}}^{\frac{N}{2}} \binom{N}{\frac{N}{2}{+}m} \left( \frac{1}{\sqrt{2}} \right)^{\frac{N}{2}{+}m}\left( \frac{1}{\sqrt{2}} \right)^{\frac{N}{2}{-}m} \ket{\frac{N}{2},m}.
\end{multline}
The state described above is known as a coherent spin state (CSS)~\cite{pezze2018quantum}. In this state, all spins are in separable states and are collectively aligned in the same direction, which we assume to be the $z-$direction. We now numerically simulate the QFI for the reduced density matrix of the spins as the system evolves from the initial state $|\psi(0)\rangle_\mathrm{CSS}{=}|\mathrm{CSS}\rangle|\alpha\rangle$. In Figs.~\ref{fig_sm_vicinity}, we illustrate such QFI for the CSS spin subsystem (orange dashed curve) as a function of the number of spins $N$ for different values of the coupling parameter $k$ and evolution times $\tau$. As the figure shows, in most cases, the QFI for the GHZ spin subsystem significantly outperforms that of the CSS spin subsystem. This remains true even in the absence of the Heisenberg limit of precision. However, for very large $N \gg 1$, the exponential factor $e^{2 k^2 N^2 (\cos \tau - 1)}$ dominates, causing the QFI for the GHZ spin subsystem to eventually vanish. Only when the prefactor $e^{2 k^2 N^2 (\cos \tau - 1)}$ becomes larger---by increasing $k$ and choosing $\tau$ far from $\tau = 2\pi$---does the QFI for the CSS spin subsystem perform better than in the GHZ case. Note that, in Fig.~\ref{fig_sm_vicinity}(d), the QFI for the CSS spin subsystem is observed to be sublinear (this is because, in most cases, accessing only the spin subsystem results in a loss of information about the gravitational acceleration $g$ and largely the information is mostly encoded in the mechanical mode). To demonstrate that the QFI for the entire spin-mechanical system, which evolves from the initial state $ |\psi(0)\rangle{=}|\mathrm{CSS}\rangle $, is linear for CSS, one can directly obtain its analytical solution as follows:
\begin{multline}
    \mathcal{Q}_\mathrm{spin}^{\mathrm{(CSS)}}(\tau){=}8{-}8\cos\tau{+}\frac{2k^2N \Gamma\left(\frac{1{+}N}{2}\right)}{\sqrt{\pi}} (\tau - \sin\tau)^2\times\\
    \left[\frac{2{+}3N}{\Gamma\left(2{+}\frac{N}{2}\right)}{+}\frac{(N{-}2)}{\Gamma\left(\frac{6{+}N}{2}\right)} N \, {}_2F_1\left(1, 2{-}\frac{N}{2}; \frac{6{+}N}{2}; -1\right)\right],
\end{multline}
where $_2F_1(a,b;c;z)$ is the hypergeometric function. In the particular case of $\tau{=}2\pi$, the QFI function reduces to:
\begin{multline}
\mathcal{Q}_\mathrm{spin}^{\mathrm{(CSS)}}(\tau{=}2\pi){=}8\pi^{3/2}k^2N \Gamma\left(\frac{1{+}N}{2}\right)\Bigg[\frac{2{+}3N}{\Gamma\left(2{+}\frac{N}{2}\right)}\\
{+}\frac{(N{-}2)N{}_2F_1(1{,}2{-}\frac{N}{2};\frac{6{+}N}{2};-1)}{\Gamma(\frac{6{-}N}{2})}\Bigg],
\end{multline}
which can be straightforwardly verified to scale linearly with $N$.

\section{Experimental feasibility}\label{sec_feasibility}

Several experimental ingredients have been achieved that could potentially be used to implement this proposal. These include: a single spin coupled parametrically to a mechanical oscillator has been demonstrated experimentally~\cite{kolkowitz2012coherent, arcizet2011single, lahaye2009nanomechanical, martinetz2020quantum}, whereas other hybrid tripartite architectures have also been realized~\cite{pirkkalainen2013hybrid, aporvari2021strong, pirkkalainen2015cavity}. On the other hand, mechanical oscillators have been cooled to their ground state~\cite{gieseler2012subkelvin, li2011millikelvin, oconnell2010quantum, cattiaux2021macroscopic}, enabling coherent state initialization. Typical mechanical frequencies of the fundamental mode range from $\omega/2\pi{\approx}50\text{MHz}$ with coupling strength $k/2\pi{\approx}0.6{-}4.6\text{MHz}$~\cite{lahaye2009nanomechanical} to $\omega/2\pi{\approx}80\text{kHz}$ with coupling strength $k/2\pi{\approx}20\text{Hz}$~\cite{kolkowitz2012coherent}. The most challenging step for the spin-mechanical proposal is the GHZ initialization of the spin subsystem. Recent theoretical proposals address this~\cite{wei2006generation, neto2017steady}, and high-fidelity 2000-atom GHZ states have been proposed~\cite{zhao2021creation}, along with few-to-14 trapped ions~\cite{friis2018observation, leibfried2004toward, monz201114qubit, roos2004control, sackett2000experimental}, 10-qubit superconducting qubits~\cite{song201710qubit}, 7-qubit GHZ states in a solid-state spin register~\cite{bradley201910qubit}, superconducting transmon qutrits~\cite{cervera2022experimental}, and three-qubit NMR systems~\cite{ji2019experimental}. Thus, current devices could feasibly implement this proposal with a small number of particles $N$.

\section{Predicted sensitivity}\label{sec_predicted}

\begin{figure}[t]
    \centering
    \includegraphics[width=\linewidth]{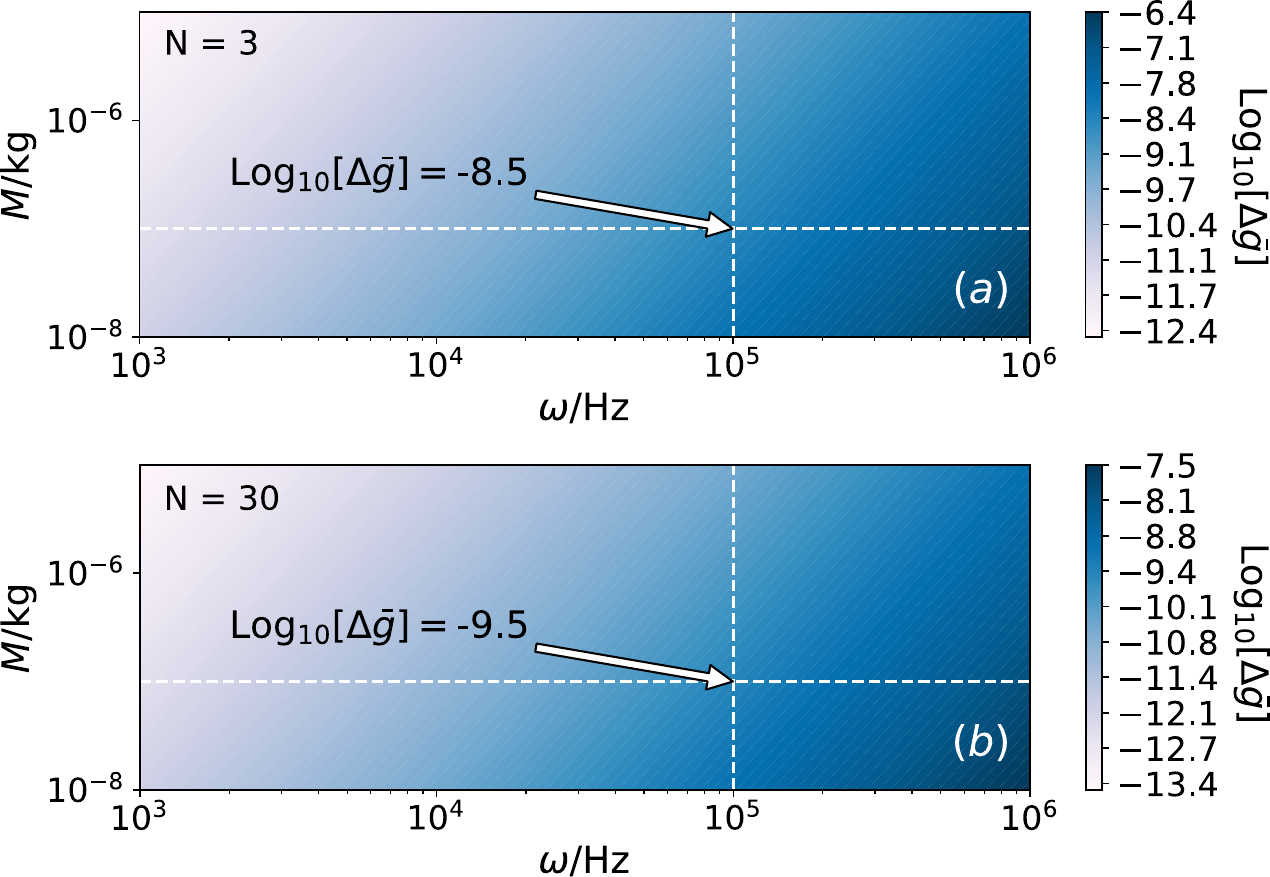}
    \caption{Absolute gravimetry sensitivity $\Delta\bar{g}$ plotted on a $\mathrm{log}_{10}$ scale as a function of $\omega$ and $M$ in SI units for various values of $N$. The arrow highlights that a moderate set of parameters can achieve predicted sensitivities of $\Delta \bar{g} \sim 10^{-9}$ m$/$s$^2$.}
    \label{sm_fig_predicted}
\end{figure}

To quantify the sensitivity of our scheme, we can evaluate $\Delta \bar{g}$ as follows~\cite{armata2017quantum}:
\begin{equation}
    \Delta \bar{g} = \frac{1}{\sqrt{\nu \mathcal{Q}_\mathrm{sm}}},
\end{equation}
Here, $\nu$ represents the number of measurement trials, $\mathcal{Q}_\mathrm{sm}$ denotes the QFI of the entire spin-mechanical system, and $\bar{g}$ is the magnitude of gravitational acceleration in the International System of Units (SI), with $[\bar{g}] = \frac{m}{s^2}$. For reference, see the Hamiltonian in Eq.~\eqref{eq_hamiltonian}. Throughout our work, we have rescaled $\bar{g}$, making it straightforward to revert to SI units by applying the chain rule: $\frac{\partial}{\partial \bar{g}} = \frac{\partial}{\partial g} \frac{\partial g}{\partial \bar{g}}$. Therefore,
\begin{equation}\mathcal{Q}_\mathrm{sm}{=}4\left(\frac{\partial g}{\partial \bar{g}}\right)^2\mathbb{R}\text{e}\left[\langle \partial_g \psi(\tau) | \partial_g \psi(\tau) \rangle{-}|\langle \partial_g \psi(\tau) | \psi(\tau) \rangle|^2\right].\end{equation}
We aim to quantify the predicted sensitivity $\Delta \bar{g}$ for a vertically oriented spin-mechanical gravimeter $\xi=0$, specifically when $\nu=10^3$, $\tau = 2\pi$ and the spin-coupling constant $k$ is of the order of unity, with $k = 1$. Under these conditions, and recalling that $\hbar g{:=}\bar{g}\sqrt{M\hbar/(2\omega^3)}\cos\xi$, the sensitivity in SI units simplifies to:
\begin{equation}
    \Delta \bar{g} \approx 10^{-19}\frac{1}{N}\sqrt{ \frac{\omega^3}{M} }.\label{sm_eq_predicted}
\end{equation}
Here, $N$ is the number of spin particles, $\omega$ is the angular frequency, and $M$ is the mass of the mechanical harmonic oscillator. Eq.~\eqref{sm_eq_predicted} presents the predicted gravimetry sensitivity for this work. Since $\omega$ and $M$ vary depending on the physical device, in Fig.~\ref{sm_fig_predicted}, we plot the predicted sensitivity $\Delta\bar{g}$ on a $\mathrm{log}_{10}$ scale as a function of $\omega$ and $M$ in SI units for different values of $N$. As highlighted by an arrow in the figure, a very moderate set of parameters could achieve predicted sensitivities of $\Delta \bar{g} \sim 10^{-9} m/s^2$. This sensitivity could be further improved by adjusting the mass and frequency. Experimentally, the sensitivity $\Delta \bar{g}$ typically ranges from $10^{-5}m/s^2$ to $10^{-10}m/s^2$, while theoretical predictions span from $10^{-7}m/s^2$ to $10^{-15}m/s^2$, see comparison table in Ref.~\cite{qvarfort2018gravimetry}. Thus, even for $N=3$ particles shown in Fig.~\ref{sm_fig_predicted}(a), this spin-mechanical gravimeter could potentially compete with other proposals, without relying on free-fall-based methodologies or the need for ground-state cooling.

\section{Conclusions}\label{sec_conclusions}

In this work, we prove that the sensitivity of gravimetry can increase quadratically with the number of spins in a conditional displacement spin-mechanical system; ultimate sensing performance known as the Heisenberg limit of precision. We show that during each mechanical cycle, the information content of the gravitational acceleration is entirely transferred to the spin subsystem. This phenomenon, resulting from the parametric interaction of the spin-mechanical probe, enables us to identify the optimal measurement basis for achieving the Heisenberg limit by only probing the spin subsystem. Note that, in general, full access to the quantum probe is required to achieve the Heisenberg limit, as probing the system locally results in a loss of information about the unknown parameters. Therefore, achieving a Heisenberg-limited gravimeter by probing only the spin subsystem, using a spin magnetization measurement basis, is a key result. Finally, we demonstrate that the proposed scheme is robust against spin-mechanical coupling anisotropies and moderate decoherence and that our work could compete with other proposals without relying on free-fall-based methodologies or the need for mechanical ground-state cooling.

\section*{Acknowledgements}

V.M. acknowledges support from the National Natural Science Foundation of China Grants No. 12374482 and No. W2432005. V.M. would like to thank George Mihailescu and Chiranjib Mukhopadhyay for their useful comments and discussions.

\appendix

\section{Spin-mechanical QFI for $N=1$ spins}\label{app_QFIsm_N1_spins}

For a single spin coupled to a mechanical harmonic oscillator, the unitary operator reads as:
\begin{multline}
\hat{U}(\tau){=}e^{{-}2ig\cos\xi \frac{k}{2}\hat{\sigma}^z[\tau{-}\sin\tau] }e^{(\frac{k}{2} \hat{\sigma}^z{-}g\cos\xi)(\eta(\tau)\hat{a}^\dagger{-} \eta(\tau)^* \hat{a})}\\
\times e^{-i\hat{a}^\dagger\hat{a}\tau},
\end{multline}
where we have used $\eta(\tau) = 1-e^{-i\tau}$ and $\omega t=\tau$. For the initial state 
\begin{equation}
    |\psi(0)\rangle_\mathrm{sm}=\left(\cos\frac{\theta}{2}|0\rangle + \sin\frac{\theta}{2}e^{-i\phi}|1\rangle\right)\otimes|\alpha\rangle,
\end{equation}
one gets:
\begin{eqnarray}
    \nonumber |\psi(\tau)\rangle &=& \cos\frac{\theta}{2} e^{\iota_+} e^{-\frac{1}{2}|\varphi_+|^2}\sum_{n=0}^\infty \frac{\varphi_+^n}{\sqrt{n!}}|0\rangle|n\rangle \\
    &+& \sin\frac{\theta}{2}e^{-i\phi} e^{\iota_-} e^{-\frac{1}{2}|\varphi_-|^2}\sum_{n'=0}^\infty \frac{\varphi_-^{n'}}{\sqrt{n'!}}|1\rangle|n'\rangle.
\end{eqnarray}
In the above, we defined $\varphi_\pm{:=}\alpha e^{-i\tau}{\pm}(\frac{k}{2}{\mp}g\cos\xi)\eta(\tau)$ and $    e^{\iota_\pm}{:=}e^{\mp igk(\tau - \sin\tau)} e^{\pm i\alpha(\frac{k}{2}\mp g\cos\xi)\sin\tau}$.

To evaluate the QFI of Eq.~\eqref{eq_QFI_all}, we compute:
\begin{multline}
 \langle \partial_g \psi(\tau) | \partial_g \psi(\tau) \rangle{=}\frac{1}{2}\cos^2\xi[4\alpha^2{-}(4 \alpha^2{+}k^2)\cos(2 \tau){+}4\\
   {-}4 k \sin\tau \{2 \alpha (\sin\tau{-}\tau) \cos\theta{+}k \tau\}{+}2 k^2 \tau^2{+}k^2{-}4 \cos\tau],
\end{multline}

and
\begin{equation}
    |\langle \partial_g \psi(\tau) | \psi(\tau) \rangle|^2 {=}\cos ^2\xi (2 \alpha \sin \tau{+}k (\tau{-}\sin\tau) \cos \theta)^2.
\end{equation}
Therefore, the QFI of the entire spin-mechanical system is:
\begin{multline}
\mathcal{Q}_\mathrm{sm}{=}\cos^2\xi(k^2 (2 \tau^2{+}1){+}k^2[{-}2(\tau{-}\sin\tau)^2\\
\cos (2 \theta){-}4 \tau \sin\tau{-}\cos (2\tau)]{-}8 \cos\tau{+}8).
\end{multline}
From the above, the optimal probe is given by $\theta = \frac{\pi}{2}$, which results in $|\psi(0)\rangle_\mathrm{sm}=\frac{1}{\sqrt{2}}\left(|0\rangle + e^{-i\phi}|1\rangle\right)|\alpha\rangle$ with an irrelevant $\phi$ angle. This leads to the optimal probe shown in the main text, $|\psi(0)\rangle_\mathrm{sm}=\ket{+}|\alpha\rangle$. Hence:
\begin{eqnarray}
\mathcal{Q}_\mathrm{sm}{=}k^2 (4 \tau^2{+}2){-}2 k^2 [4 \tau\sin\tau{+}\cos (2\tau)]{-}8 \cos\tau{+}8.
\end{eqnarray}
The above QFI of the entire system is the one presented in the main text; see Eq.~\eqref{eq_qsm1}.

\section{Spin-mechanical QFI for $N=2$ spins}\label{app_QFIsm_N2_spins}

Similarly to the single spin case shown above, we first explicitly derive the unitary temporal operator for two spins as follows:
\begin{multline}
    \hat{U}(\tau) = e^{ i (2\frac{k_1}{2}\frac{k_2}{2} \hat{\sigma}_1^z\hat{\sigma}_2^z - 2 \frac{k_1}{2} g\cos\xi\hat{\sigma}_1^z - 2 \frac{k_2}{2}g\cos\xi\hat{\sigma}_2^z)[\tau - \sin\tau] }\\
    e^{\left(  \frac{k_1}{2} \hat{\sigma}_1^z + \frac{k_2}{2} \hat{\sigma}_2^z - g\cos\xi\right)(\eta(\tau) a^\dagger - \eta(\tau)^* a)} e^{-i a^\dagger a \tau}.
\end{multline}
We evolve the state from a general two spins state in the computational basis with a coherent state for the field:
\begin{equation}
    |\psi(0)\rangle_\mathrm{sm}^{(2)}{=}(R_1|01\rangle{+}R_2|00\rangle{+}R_3|10\rangle{+}R_4|11\rangle)|\alpha\rangle,
\end{equation}
where $R_j{=}r_je^{-i\phi_j}$ and $\sum\limits_{j{=}1}^4|R_j|^2{=}1$. The QFI for the two spins case is:
\begin{multline}
    \mathcal{Q}_\mathrm{sm}^{(2)}{=}{-}8 \cos ^2\xi(2\tau^2(k_1^2 (r_3^2{+}r_4^2{-}1) (r_3^2{+}r_4^2){-}2 k_1 k_2 r_3^2(r_2^2\\
    {+}r_3^2{-}1){+}k_2^2 (r_2^2{+}r_3^2-1)(r_2^2{+}r_3^2)){+}2\sin\tau(\sin\tau{-}2\tau)[k_1^2 (r_3^2\\
    {+}r_4^2{-}1) (r_3^2{+}r_4^2){-}2 k_1 k_2 r_4^2 (r_2^2{+}r_3^2){-}2 k_1 k_2 r_3^2 (r_2^2{+}r_3^2{-}1)\\
    +k_2^2 (r_2^2{+}r_3^2{-}1) (r_2^2{+}r_3^2)]{+}\cos\tau{-}1{-}2 k_1 k_2 r_4^2 (r_2^2{+}r_3^2)).
\end{multline}
For the specific case of $k_1=k_2=k$, and maximized over $r_i's$ ($i=2,3,4$) amplitudes, one gets:
\begin{eqnarray}
   \nonumber \mathcal{Q}_\mathrm{sm}^{(2)}&=&8 \left(2 k^2 \tau^2{+}2 k^2 \sin\tau (\sin\tau{-}2\tau){-}\cos\tau{+}1\right),\\
   &\propto& 4k^2(\tau{-}\sin\tau)^2{-}2\cos\tau,
\end{eqnarray}
which results in the QFI of the entire spin-mechanical system shown in the main text; see Eq.~\eqref{eq_qsm2}.

\section{Spin-mechanical QFI for $N$ spins: Greenberger-Horne-Zeilinger state}\label{app_QFIsm_GHZ}

As discussed in the main text, for the particular case of equal spin-mechanical coupling strengths $k_i = k_j$ for all $i$ and $j$, one can employ the collective spin operators, which simplifies the operator $\hat{Z}(\bm{k}, g) = k\hat{S}_z - g$ of the Hamiltonian shown in Eq.~\eqref{eq_hamiltonian_scaled}. With this choice, the unitary temporal operator reduces to:
\begin{multline}
    \hat{U}(\tau){=}e^{i(k\hat{S}_z{-}g\cos\xi)^2(\tau{-}\sin\tau)}e^{(k\hat{S}_z{-}g\cos\xi)(\eta(\tau)\hat{a}^\dagger{-}\eta(\tau)^*\hat{a})}\\
    \times e^{-i\tau\hat{a}^\dagger\hat{a}}.\label{sm_unitary_operator}
\end{multline}
By considering the initial state composed of the Greenberger-Horne-Zeilinger state and a coherent state with real amplitude $\alpha$:
\begin{equation}
\ket{\psi(0)}_\mathrm{sm}^{\text{(GHZ)}}{=}\frac{1}{\sqrt{2}}\left(\ket{\frac{N}{2}{,}\frac{N}{2}}{+}\ket{\frac{N}{2}{,}-\frac{N}{2}}\right)\ket{\alpha},
\end{equation}
one gets the evolved wave function as:
\begin{multline}
\ket{\psi(\tau)}_\mathrm{sm}^{\text{(GHZ)}}{=}\frac{1}{\sqrt{2}}\Big[e^{\iota_{+,N}}\ket{\frac{N}{2}{,}\frac{N}{2}}\ket{\varphi_{+,N}}\\
{+}e^{\iota_{-,N}}\ket{\frac{N}{2}{,}-\frac{N}{2}}\ket{\varphi_{-,N}}\Big],\label{sm_eq_psitghz}
\end{multline}
where $e^{\iota_{\pm, N}}{:=}e^{ i (k\frac{N}{2}{\mp}g\cos\xi)^2(\tau{-}\sin\tau) } e^{\pm i \alpha\sin\tau(k\frac{N}{2}\mp g \cos\xi)}$ and $\varphi_{\pm,N}{:=}\alpha e^{-i\tau}{\pm}\left(k\frac{N}{2}{\mp}g\cos\xi\right)\eta(\tau)$. From the above, it is then straightforward to compute the QFI for this case:
\begin{multline}
    \mathcal{Q}_\mathrm{sm}^{(\text{GHZ})}{=}2 \cos^2\xi(k^2N^2 (2 \tau^2+1)\\
    -k^2 N^2 [4 \tau \sin\tau+\cos (2\tau)]-4 \cos\tau+4),
\end{multline}
which corresponds to the QFI shown in the main text, see Eq.~\eqref{eq_GHZ_Heisenberg_Limit_sm}.

\section{Gravimetry without the necessity of mechanical ground state cooling}\label{app_groundstate}

A simple yet highly significant observation is that the spin-mechanical system becomes completely disentangled at times that are multiples of $2\pi$. This is an intrinsic feature of the system's conditional displacement interaction, $\hat{Z}(\boldsymbol{k},g)$ in the main text. In what follows, we demonstrate that this disentanglement occurs even when the mechanical oscillator is initialized in a thermal state at an arbitrary temperature $T$, encoded in the average phonon excitation $\bar{n}=(\mathrm{exp}[\hbar\omega/(k_BT)] - 1)^{-1}$, where $k_B$ is the Boltzmann constant. This is critically important for experimental feasibility, as other approaches, such as free-fall atomic interferometry with cold atoms, require the atomic ensemble to be cooled to cryogenic temperatures. To show that ground state cooling is not necessary, we evolve the following initial state (we write $\ket{\frac{N}{2}{,}\pm\frac{N}{2}}=\ket{\pm\frac{N}{2}}$ for simplicity):
\begin{multline}
    \rho(0) = \frac{1}{2}\left(\ket{\frac{N}{2}}{+}\ket{-\frac{N}{2}}\right) \left(\bra{\frac{N}{2}}{+}\bra{-\frac{N}{2}}\right)\\
    \otimes\frac{1}{\pi\bar{n}}\int d^2\alpha |\alpha\rangle \langle \alpha|e^{-\frac{|\alpha|^2}{2}}.
\end{multline}
Using the unitary temporal operator in Eq.~\eqref{sm_unitary_operator}, the spin-mechanical state evolves as:
\begin{multline}
    \rho(\tau)=\frac{1}{2\pi\bar{n}}\int d^2\alpha e^{-\frac{|\alpha|^2}{2}}\Bigg[ \ket{\varphi_{+,N}}\bra{\varphi_{+,N}}\otimes\ket{\frac{N}{2}}\bra{\frac{N}{2}}\\
    +\ket{\varphi_{-,N}}\bra{\varphi_{-,N}}\otimes\ket{-\frac{N}{2}}\bra{-\frac{N}{2}}\\
    +\Bigg(e^{\iota_{+,N}}e^{-\iota_{-,N}}\ket{\frac{N}{2}}\bra{-\frac{N}{2}}\otimes\ket{\varphi_{+,N}}\bra{\varphi_{-,N}}{+}h.c\Bigg)\Bigg].
\end{multline}
From the above expression, it is evident that at times $\tau$ that are multiples of $2\pi$, $\eta(\tau = 2\pi) = 0$, and thus:
\begin{multline}
    \rho(\tau{=}2\pi){=}\frac{1}{2}\Bigg(\ket{\frac{N}{2}}{+}e^{4\pi i kNg\cos\xi}\ket{{-}\frac{N}{2}}\Bigg)\Bigg(\bra{\frac{N}{2}}\\
    + e^{-4\pi ik N g\cos\xi}\bra{-\frac{N}{2}}\Bigg)\otimes\frac{1}{\pi\bar{n}}\int d^2\alpha |\alpha\rangle \langle \alpha|e^{-\frac{|\alpha|^2}{2}}.
\end{multline}
Finally, it is clear that the gravitational acceleration is entirely transferred to the spin subsystem as a relative phase between the states $\{\ket{\frac{N}{2}, \frac{N}{2}}, \ket{\frac{N}{2}, -\frac{N}{2}}\}$, in the same way it would be if the mechanical oscillator evolved from an initial coherent state. This demonstrates that it is not necessary to cool the mechanical oscillator to its ground state, as long as the system is allowed to evolve for times that are multiples of $2\pi$.

\section{Spin QFI for $N$ particles: reduced density matrix of the spin subsystem}\label{app_QFIs_Nspins}

As discussed in the main text, we can investigate the precision limits when there is partial accessibility to the system. Specifically, to determine the sensing capability of the spin subsystem, we compute its reduced density matrix as follows:
\begin{equation}
    \rho_\mathrm{spin} = \rho_\mathrm{spin}(g) =  \text{Tr}_\text{field}\left[ (\ket{\psi(\tau)}\bra{\psi(\tau)})_\mathrm{sm}^{(\text{GHZ})}  \right],
\end{equation}
where the state $\ket{\psi(\tau)}_\mathrm{sm}^{(\text{GHZ})}$ has been derived in Eq.~\eqref{sm_eq_psitghz}. To compute the QFI for density matrices, we use:
\begin{equation}
\mathcal{Q}_\mathrm{sm} = 2 \sum_{n,m} \frac{|\langle \lambda_m |\partial_g \rho_\mathrm{spin}| \lambda_n \rangle|^2}{\lambda_n + \lambda_m}, \hspace{0.5cm} \lambda_n + \lambda_m \neq 0,
\end{equation}
where $\rho_\mathrm{spin} = \sum_i \lambda_i |\lambda_i\rangle\langle\lambda_i|$ is represented in its spectral decomposition, with $\lambda_i$ and $|\lambda_i\rangle$ being the $i$th eigenvalue and eigenvector of $\rho_\mathrm{spin}$, respectively. Hence, let us gather the necessary ingredients to evaluate the QFI. The reduced density matrix is:
\begin{equation}
\rho_\mathrm{spin} = \frac{1}{2}\left(
\begin{array}{cc}
 1 & \rho_\mathrm{spin}^{(1,2)}\\
  \rho_\mathrm{spin}^{(2,1)} & 1 \\
\end{array}
\right),
\end{equation}
where
\begin{eqnarray*}
\rho_\mathrm{spin}^{(1,2)}&=&e^{k^2 L^2 (\cos\tau-1)} e^{2ikN(\alpha\sin\tau+g\cos\xi)(\sin\tau - \tau)},\\
\rho_\mathrm{spin}^{(2,1)}&=&e^{k^2 L^2 (\cos\tau-1)} e^{-2ikN(\alpha\sin\tau+g\cos\xi)(\sin\tau - \tau)},
\end{eqnarray*}
with eigenvalues:
\begin{eqnarray}
    \lambda_1 &=& \frac{1}{2}\left( 1 - e^{k^2N^2(\cos\tau - 1} \right), \\
    \lambda_2 &=& \frac{1}{2}\left( 1 + e^{k^2N^2(\cos\tau - 1} \right),
\end{eqnarray}
and eigenvectors:
\begin{multline}
    |\lambda_1\rangle=\frac{1}{\sqrt{2}}\ket{{-}\frac{N}{2}}{-}\frac{e^{2ikN(\alpha\sin\tau{+}g\cos\xi(\sin\tau{-}\tau))}}{\sqrt{2}}\ket{\frac{N}{2}},\\
    |\lambda_2\rangle=\frac{1}{\sqrt{2}}\ket{{-}\frac{N}{2}}{+}\frac{e^{2ikN(\alpha\sin\tau{+}g\cos\xi(\sin\tau{-}\tau))}\ket{\frac{N}{2}}}{\sqrt{2}}.
\end{multline}
From which one can easily derive the expression shown in the main text; see Eq.~\eqref{eq_Q_spin}:
\begin{eqnarray}
\nonumber \mathcal{Q}_\mathrm{sm} &=& 2 \sum_{n,m} \frac{|\langle \lambda_m |\partial_g \rho_\mathrm{spin}| \lambda_n \rangle|^2}{\lambda_n + \lambda_m}\\
&=& 4e^{2 k^2N^2(\cos\tau{-}1)}k^2N^2(\tau{-}\sin\tau)^2.
\end{eqnarray}

\section{CFI for the spin subsystem: analytical form}\label{app_CFIs_analytical}

Here, we present the analytical form of the CFI function for the reduced density matrix of the spin subsystem $\mathcal{F}_\mathrm{spin}$. To achieve this, we directly evaluate the conditional probability distribution $p(\Upsilon(\Theta, \Phi) | g)$ as follows:
\begin{multline}
    p(\Upsilon(\Theta{,}\Phi)|g){=}\bra{\psi(\tau)}\hat{\Pi}_\Upsilon\ket{\psi(\tau)}{=}\frac{1}{2}\Big(1{+}e^{k^2N^2(\cos\tau{-}1)}\\
    \times \sin\theta\cos[\phi{+}2gkN\tau\cos\xi{-}2kN[\alpha{+}g\cos\xi]\sin\tau]\Big).
\end{multline}
With the above expression, it is straightforward to evaluate the CFI as follows:
\begin{equation}
    \mathcal{F}_\mathrm{spin} = \frac{\mathcal{F}_\mathrm{spin}^\mathrm{num.}}{\mathcal{F}_\mathrm{spin}^\mathrm{den.}},
\end{equation}
where
\begin{multline}
\mathcal{F}_\mathrm{spin}^\mathrm{num.}{=}-4e^{2k^2N^2(\cos\tau{-}1)}k^2N^2\cos^2\xi(t{-}\sin\tau)^2 \sin^2\theta\\
\times \sin^2\left(\phi{+}2gkN\tau\cos\xi{-}2kN(\alpha{+}g\cos\xi)\sin\tau \right)\\
    \mathcal{F}_\mathrm{spin}^\mathrm{den.}{=}\cos^2\left(\phi{+}2gkN\tau \cos\xi{-}2kN(\alpha{+}g\cos\xi)\sin\tau\right)\\
    \times e^{2k^2N^2(\cos\tau{-}1)} \sin^2\theta {-}1.
\end{multline}

As the above expression shows, the comparison between $\mathcal{F}_\mathrm{spin}$ and $\mathcal{Q}_\mathrm{spin}$ must be made with respect to the product $kN$. This is because both the quantum and CFI functions inherently depend on the term $kN$.

\bibliography{Gravimetry}

\end{document}